\documentclass[manuscript,screen,acmsmall,nonacm]{acmart}
\acmJournal{TEAC}
\AtBeginDocument{%
  }

\setcopyright{acmlicensed}
\copyrightyear{2026}
\acmYear{2026}
\acmDOI{XXXXXXX.XXXXXXX}

\thanks{A preliminary version of this paper appeared in the Proceedings of SAGT 2025 \citep{gspsagt} and the full version in the ACM Transactions on Economics and Computation \citep{gspfull}.}

\usepackage{pgfplots}
\usepackage{subcaption}
\pgfplotsset{compat=1.18}

\definecolor{acmDarkBlue}{RGB}{0,114,178}
\definecolor{acmGreen}{RGB}{0,158,115}
\definecolor{acmPink}{RGB}{204,121,167}
\definecolor{acmOrange}{RGB}{213,94,0}
\definecolor{acmYellow}{RGB}{240,228,66}
\definecolor{acmLightBlue}{RGB}{86,180,233}

\pgfplotsset{
    cycle list={
        {acmDarkBlue, mark=*},
        {acmGreen, mark=square*},
        {acmPink, mark=triangle*},
        {acmOrange, mark=diamond*},
        {acmYellow, mark=o},
        {acmLightBlue, mark=star}
    }
}
\usepackage{mathtools}
\usepackage{thmtools}
\usepackage{tikz}
\usepackage{algorithm}
\usepackage{algpseudocode}
\usepackage{hyperref}
\usepackage{outlines}
\usepackage{url}

\newcommand*\circled[1]{\tikz[baseline=(char.base)]{
            \node[shape=circle,draw,inner sep=1.2pt] (char) {#1};}}
\newcommand*\dottedcircled[1]{\tikz[baseline=(char.base)]{
            \node[shape=circle,draw,dotted,inner sep=1.2pt] (char) {#1};}}

\usepackage{xcolor, soul}

\usepackage{caption}

\newtheorem{definition}{Definition}[section]
\newtheorem{theorem}[definition]{Theorem}
\newtheorem{corollary}[definition]{Corollary}
\newtheorem{lemma}[definition]{Lemma}

\newtheorem{observation}[definition]{Observation}
\newtheorem{example}[definition]{Example}

\usepackage{enumitem}



\begin{document}

\title{Unsolvability and Beyond in Many-to-Many Non-bipartite Stable Matching}

\author{Frederik Glitzner}
\email{f.glitzner.1@research.gla.ac.uk}
\orcid{0009-0002-2815-6368}
\affiliation{
 \institution{School of Computing Science, University of Glasgow}
  \city{Glasgow}
  \country{United Kingdom}
}

\author{David Manlove}
\email{david.manlove@glasgow.ac.uk}
\orcid{0000-0001-6754-7308}
\affiliation{
  \institution{School of Computing Science, University of Glasgow}
  \city{Glasgow}
  \country{United Kingdom}
}

\renewcommand{\shortauthors}{F. Glitzner and D. Manlove}

\begin{abstract}
    We study the {\sc Stable Fixtures} problem, a many-to-many generalisation of the classical non-bipartite {\sc Stable Roommates} matching problem. Building on the foundational work of Tan on stable partitions, we extend his results to this significantly more general setting and develop a rich framework for understanding stable structures. Our main contribution, the notion of a \emph{generalised stable partition (GSP)}, not only characterises the solution space but also serves as a versatile tool for ordinal preference systems with capacity constraints. 
    
    We show that a GSP can be computed efficiently and can provide an elegant representation of key aspects of a preference system. Leveraging a connection to stable half-matchings, we also establish an analogous Rural Hospitals Theorem for stable half-matchings and GSPs, and connect our results to recent work on near-feasible matchings, providing a simpler algorithm and tighter analysis. 
    
    Our work also addresses the computational challenges of finding optimal stable half-matchings and GSPs, presenting a flexible integer linear programming model for various objectives. Beyond theoretical insights, we conduct the first empirical analysis of random {\sc Stable Fixtures} instances. Our work unifies and extends classical and recent perspectives on stability in non-bipartite stable matching and establishes new tools and techniques for stable matchings and their applications.
\end{abstract}

\ccsdesc[500]{Mathematics of computing~Combinatorial algorithms}
\ccsdesc[500]{Mathematics of computing~Matchings and factors}
\ccsdesc[500]{Theory of computation~Algorithmic game theory}

\keywords{Stable Matching, Stable Fixtures, Stable Partition, Generalised Stable Partition}


\maketitle

\section{Introduction}

Stable matching theory is a central topic in algorithmic game theory (see, for example, Manlove \citep{matchup} for an extensive overview of algorithmic questions and advances), attracting attention from researchers in economics, computer science, and operations research. One of the foundational problems in this area is the {\sc Stable Roommates} problem ({\sc sr}), first proposed and studied by Gale and Shapley \citep{galeShapley}, where agents seek to be matched in pairs without creating incentives for deviation. Note that {\sc sr} is a non-bipartite generalisation of the classical bipartite {\sc Stable Marriage} problem, which deals with two disjoint sets of agents. However, in contrast to its bipartite restriction, {\sc sr} suffers from a major limitation: stable matchings do not always exist (see \citet{gusfield89} for an extensive overview of structural insights). This lack of guaranteed feasibility has motivated researchers to look for structural characterisations and alternative solution concepts.

A significant advance \citep{matchup} came in the form of \emph{stable partitions} -- a combinatorial structure introduced by Tan \citep{tan91_1} as a relaxation of a stable matching that always exists and, crucially, tightly characterises the structure of {\sc sr} instances and certifies the \emph{unsolvability} (the non-existence of a stable matching) of such instances succinctly. Stable partitions also correspond naturally to stable half-matchings \citep{biro08}, where agents may split their capacity among two partners, thereby allowing for a relaxed, yet stable, assignment.

A natural generalisation of {\sc sr} to a many-to-many setting is the {\sc Stable Fixtures} problem ({\sc sf}), first studied by Irving and Scott \citep{Irving2007}, where agents have integer capacities and may be matched to multiple other agents. Of course, {\sc sf} is again a non-bipartite generalisation of the classical bipartite counterparts of the problem, such as the {\sc Hospitals/Residents} problem (in which one side of the market has capacity 1 and the other side can have arbitrary capacities) \citep{matchup}. Furthermore, Csáji et al. \citep{couples24} recently made an interesting connection here by using the {\sc sf} model to establish new results for instances (with certain properties) of the {\sc Hospitals/Residents with Couples} problem (in which residents can apply to hospitals in pairs). The {\sc sf} model, and related non-bipartite matching models, also have applications in engineering, such as in peer-to-peer network design (see, for example, \citet{lebedev2007matching} and \citet{Mathieu2010}). While {\sc sf} inherits many of the complexities of {\sc sr}, it also introduces new combinatorial phenomena and richer matching structures. Despite the increased generality, many concepts carry over, and the existence of a stable half-matching is guaranteed \citep{Fleiner08}, offering a promising framework to extend the notion of stable partitions to the {\sc sf} domain. 

This work, continuing a line of research by \citet{glitzner24sagt,glitzner2024structural}, proposes and investigates such a generalisation: we introduce the concept of \emph{generalised stable partitions} in the {\sc sf} setting, which formalise and characterise stable structures in the many-to-many context. This perspective enables us to unify and generalise several well-studied topics and influential results from stable matching theory, while also shedding light on unsolvability in {\sc sf} instances through a new lens. Beyond deep insights into the structure of solvable and unsolvable {\sc sf} instances, our work provides the necessary tools to investigate other solution concepts and problems related to active areas of research, such as near-feasibility \citep{gergely25}, almost-stability \citep{chen17}, and control \citep{Chen2025}, and to establish new results in other problem models, such as {\sc Hospitals/Residents with Couples}.

\subsection{Contributions}

We provide new structural and algorithmic insights into problem instances and stable matchings in a many-to-many non-bipartite setting, with the following main contributions:

\begin{itemize}
    \item We generalise Tan's concept of \emph{stable partitions} \citep{tan91_1} to the {\sc Stable Fixtures} setting by introducing two equivalent definitions of \emph{generalised stable partitions} (GSPs).
    \item We show that GSPs share key structural properties with their classical counterparts: they always exist, succinctly certify the unsolvability of an instance through the presence of invariant odd-length cycles (of length $\geq 3$), and can be turned into \emph{reduced GSPs} efficiently. Furthermore, we establish that reduced GSPs of solvable instances correspond bijectively to stable matchings. 
    \item We establish a formal correspondence between GSPs and \emph{stable half-matchings} \citep{Fleiner02,Fleiner08,Fleiner2010}, and use this connection to show that a GSP can be computed in $O(n^3)$ time, given a {\sc Stable Fixtures} instance with $n$ agents. 
    \item We demonstrate how GSPs enable improved results for active research problems. In particular, we provide a simple algorithm with tight structural and complexity guarantees for the near-feasible stable matching problem, which improves upon a recent result due to Csáji \citep{gergely25}.
    \item We note the intractability of several variants of the optimal stable half-matching problem \citep{glitzner24sagt} and propose a general integer linear programming (ILP) framework that can optimise for multiple objectives. Furthermore, we present an empirical analysis that shows that solvability rates and structural features (e.g., number of odd cycles) behave in somewhat surprising ways for different choices of capacity functions and that a higher capacity does not necessarily guarantee a higher likelihood of the instance being solvable.
\end{itemize}

\subsection{Structure of the Paper}

First, we will present some related work and formal definitions in Section \ref{sec:background}. Section \ref{sec:definingGSP} gives two definitions of GSPs that are fundamentally different in spirit but turn out to be equivalent. Section \ref{sec:algorithms} establishes the correspondence between GSPs and stable half-matchings and leverages this result to give a polynomial-time algorithm to compute GSPs, highlighting why natural incremental algorithms can fail. Subsequently, Section \ref{sec:structure} gives deep structural results for GSPs and, later, for near-feasible solvable instances and matchings. Section \ref{sec:exp} provides integer linear programs and experimental results for random {\sc sf} instances. Finally, Section \ref{sec:conclusion} gives a short summary of the results and proposes some future research directions.

\section{Background}
\label{sec:background}

In this section, we formally define the {\sc Stable Roommates} and {\sc Stable Fixtures} problems, and survey related work in more depth.

\subsection{Stable Roommates and Stable Partitions}

We start by defining the fundamental {\sc Stable Roommates} problem.

\begin{definition}[{\sc sr} Problem]
    Let $I=(A,\succ)$ be an {\sc sr} \emph{instance} where $A=\{a_1,a_2,\dots,a_n\}$, also denoted by $A(I)$, is a set of $n\in\mathbb{N}$ agents, and $\succ$ is a tuple of strict linear preference rankings $\succ_i$ of agents $a_i$ over all other agents $a_j\in A\setminus\{a_i\}$.

    A \emph{matching} $M$ of $I$ is a set of unordered pairs of agents in $A$ such that no agent is contained in more than one pair. We denote the partner of agent $a_i$ in $M$ by $M(a_i)$, and this can be empty.

    A pair $\{a_i,a_j\}$ of agents is \emph{blocking} in $M$ if $a_j\succ_i M(a_i)$ and $a_i\succ_j M(a_j)$. If $M$ admits no blocking pairs, then we refer to $M$ as \emph{stable}. If $I$ admits at least one stable matching, then we refer to $I$ as \emph{solvable}; otherwise, $I$ is referred to as \emph{unsolvable}.
\end{definition}

Note that throughout this paper, we will generally assume \emph{complete preferences}, i.e., every agent pair is considered acceptable (but it has been shown that most results carry over to the setting with \emph{incomplete preferences} \citep{gusfield89}, and we may take advantage of this implicitly).

The solvability of $I$ can be determined in $O(n^2)$ time using Irving's algorithm \citep{irving_sr}. It is well known that a stable matching is unlikely to exist when the number of agents is large \citep{mertens05}, although there are also families of instances that admit exponentially many \citep{gusfield89}. Therefore, Gusfield and Irving \citep{gusfield89} asked whether it is possible to provide a succinct certificate for the unsolvability of an {\sc sr} instance. The question was answered positively by Tan \citep{tan91_1}, when the author defined a combinatorial structure known as a \emph{stable partition}, although the structure's significance extends beyond its use as a witness of unsolvability. Apart from being a useful structural tool to study the problem, every stable partition corresponds to a \emph{stable half-matching} \citep{biro08}, in which agents are assigned either one other agent with value 1 (a full match) or two different agents each with value 0.5 (two half-matches). In this structure, agents are in a stable state in the sense that no two agents would rather decrease their value with one of their assigned agents in favor of increasing their intensity with each other. Tan showed that a stable partition always exists, and it has since been observed that the stable half-matching associated with a stable partition can be motivated by, for example, sports tournament scheduling and time-sharing applications \citep{biro08}. A stable partition is defined formally as follows. 

\begin{definition}[Stable Partition]
\label{def:sp}
    Let $I=(A,\succ)$ be an {\sc sr} instance. Then $\Pi$ is a \emph{stable partition} if it is a permutation of $A$ and  
    \begin{enumerate}[leftmargin=3.5em]
        \item[(T1)] $\forall a_i \in A$ we have $\Pi(a_i) \succeq_i \Pi^{-1}(a_i)$, and
        \item[(T2)] $\nexists \;a_i, a_j \in A$ with $a_i\neq a_j$ such that $a_j \succ_i \Pi^{-1}(a_i)$ and $a_i \succ_j \Pi^{-1}(a_j)$,
    \end{enumerate}
    where $\Pi(a_i) \succeq_i \Pi^{-1}(a_i)$ means that either $a_i$'s successor in $\Pi$ is equal to its predecessor, or the successor has a better rank in the preference list of $a_i$.
\end{definition}

Tan \citep{tan91_1,tan91_2} established the following crucial properties of a stable partition.

\begin{theorem}[\citep{tan91_1,tan91_2}]
\label{thm:tan}
    The following properties hold for any {\sc sr} instance $I$:
    \begin{itemize}
        \item Any two stable partitions $\Pi_a, \Pi_b$ of $I$ contain exactly the same odd cycles.
        \item $I$ admits a stable matching if and only if no stable partition of $I$ contains an odd cycle (of length at least 3).
    \end{itemize}
\end{theorem}

Note that if an {\sc sr} instance $I$ admits a stable matching $M=\{\{a_{i_1}, a_{i_2}\}, \dots, \{a_{i_{2k-1}}, a_{i_{2k}}\}\}$ consisting of $k$ pairs, then it can be represented equivalently via its induced collection of transpositions $\Pi=(a_{i_1}$ $a_{i_2})\dots (a_{i_{2k-1}}$ $a_{i_{2k}})$, which is a stable partition of $I$.

\begin{example}
In Table \ref{table:solvable}, agent $a_1$ ranks the agents $a_2$, $a_5$, $a_3$, etc., in linear order (i.e., the \emph{rank} of $a_2$ in $a_1$'s preference list is 1, of $a_5$ is 2, etc.). The problem instance admits the stable partition $\Pi=(a_1 \; a_2)(a_3 \; a_4)(a_5 \; a_6)$ (that also corresponds to a stable matching) with predecessors and successors (which are equal because $\Pi$ consists only of transpositions) indicated in boxes, the stability of which can be easily verified by hand by looking at all agents that are not matched to their first choice and seeing that no agent preferable to their partner would rather be matched to them than their current partner. On the other hand, Table \ref{table:unsolvable} shows an example instance that does not admit any stable matching, succinctly certified by its stable partition $\Pi=(a_1 \; a_2 \; a_3)(a_4 \; a_5 \; a_6)$.

\begin{table}[!htb]
\centering
\begin{minipage}{.47\linewidth}
\centering
\caption{A solvable {\sc sr} instance with a stable matching (or an equivalent stable partition consisting of transpositions only) indicated using boxes.}
    \begin{tabular}{ c | c c c c c }
    $a_i$ & pref \\\hline
    $a_1$ & $\boxed{a_2}$ & $a_5$ &  $a_3$ & $a_4$ & $a_6$ \\
    $a_2$ & $a_5$ & $a_3$ & $\boxed{a_1}$ & $a_6$ & $a_4$ \\
    $a_3$ & $\boxed{a_4}$ & $a_2$ & $a_6$ & $a_1$ & $a_5$ \\
    $a_4$ & $a_1$ & $a_2$ & $a_5$ & $\boxed{a_3}$ & $a_6$ \\
    $a_5$ & $\boxed{a_6}$ & $a_2$ & $a_1$ & $a_4$ & $a_3$ \\
    $a_6$ & $\boxed{a_5}$ & $a_3$ & $a_2$ & $a_4$ & $a_1$
    \end{tabular}
\label{table:solvable}
\end{minipage}%
\hfill
\begin{minipage}{.47\linewidth}
\centering
\caption{An unsolvable {\sc sr} instance with a stable partition indicated using broken and solid circles for successors and predecessors, respectively.}
    \begin{tabular}{ c | c c c c c }
    $a_i$ & pref \\\hline
    $a_1$ & \circled{$a_2$} & $a_5$ & \dottedcircled{$a_3$} & $a_4$ & $a_6$ \\
    $a_2$ & $a_4$ & \circled{$a_3$} & \dottedcircled{$a_1$} & $a_6$ & $a_5$ \\
    $a_3$ & $a_5$ & $a_4$ & \circled{$a_1$} & \dottedcircled{$a_2$} & $a_6$ \\
    $a_4$ & $a_1$ & \circled{$a_5$} & \dottedcircled{$a_6$} & $a_2$ & $a_3$ \\
    $a_5$ & \circled{$a_6$} & $a_2$ & \dottedcircled{$a_4$} & $a_1$ & $a_3$ \\
    $a_6$ & $a_1$ & $a_2$ & $a_3$ & \circled{$a_4$} & \dottedcircled{$a_5$} 
    \end{tabular}
\label{table:unsolvable}
\end{minipage} 
\end{table}
\end{example}

\citet{tan91_2} also showed that finding the largest solvable sub-instance (in the sense of the smallest set of agents to delete from the instance), called the \emph{maximum stable matching} problem, is naturally solvable using stable partitions by deleting exactly one arbitrary agent from each cycle of odd length found in the instance's stable partition. Recent work by \citet{glitzner24sagt,glitzner2024structural} focused on the structure and algorithmic aspects of stable partitions, as well as the complexity of finding ``optimal'' stable partitions (which can equally be thought of as optimal stable half-matchings).

\subsection{Stable Fixtures and Other Generalisations}

Irving and Scott \citep{Irving2007} introduced the {\sc Stable Fixtures} problem ({\sc sf}), a many-to-many extension of {\sc sr} in which each agent has an integral capacity indicating the maximum number of agents that they can be paired with, and showed that Irving's algorithm can be extended to find a stable matching or determine that none exist in $O(n^2)$ time \citep{Irving2007}. Interesting phenomena in {\sc sf} that do not occur in {\sc sr} were illustrated by \citet[Section 4.8.4]{matchup}.\footnote{For example, suppose that $I$ is an {\sc sf} instance with $n=2k$ agents for some $k\geq 1$. Preferences are already assumed to be complete, and let us assume furthermore that all agents have the same capacity $c$ for some $1\leq c\leq 2k-1$ (we later refer to such instances as instances with \emph{uniform capacities}). If $c=1$ then $I$ corresponds to an {\sc sr} instance and every stable matching must have size $k$. However, when $c>1$, it is not necessarily true that stable matchings are of size $ck$ as one might initially expect.} Problem instances and our stability criteria are defined formally as follows.

\begin{definition}[{\sc sf} Instance]
    Let $I=(A,\succ, c)$ be an {\sc sf} \emph{instance} where $A=\{ a_1, a_2, \dots, a_n \}$, also denoted $A(I)$, is a set of $n\in \mathbb{N}$ agents, $\succ$ is a tuple of strict linear \emph{preference rankings} $\succ_i$ of agents $a_i$ over all other agents $a_j\in A\setminus\{a_i\}$, and $c$ is a tuple of integral upper \emph{capacities} $c_i$ of agents $a_i$ giving the maximum number of agents that $a_i$ can be assigned to (where naturally $1\leq c_i\leq n-1$).
\end{definition}

\begin{definition}[Stability]
    Let $I=(A,\succ, c)$ be an {\sc sf} instance. A \emph{matching} $M$ is a set of unordered pairs of agents such that no agent $a_i$ is contained in more than $c_i$ pairs. Let worst$_i(M(a_i))$ denote the worst agent assigned to $a_i$ in $M$ (according to $\succ_i$) and $\vert M(a_i)\vert$ denote the number of agents assigned to $a_i$ in $M$. Then a \emph{blocking pair} of $M$ is a pair of distinct agents $a_i, a_j\in A$ such that 
    \begin{enumerate}[label=(\roman*),leftmargin=3.5em]
        \item $\{a_i,a_j\}\notin M$, and
        \item $\vert M(a_j)\vert< c_j$ or $a_i\succ_j$ worst$_j(M(a_j))$, and 
        \item $\vert M(a_i)\vert<c_i$ or $a_j \succ_i$ worst$_i(M(a_i))$.
    \end{enumerate}
    If $M$ does not admit any blocking pair, then it is called \emph{stable}.
\end{definition}

Henceforth, suppose that agents rank themselves last (i.e., any agent would rather be matched to anyone else than to be matched to itself, which corresponds to being \emph{unmatched} or forming a \emph{fixed-point}), as in the case in the study of stable partitions in {\sc sr}.

\begin{example}
In Table \ref{table:solvablesf}, $a_1$ has capacity 2 and ranks the agents $a_2$, $a_3$, $a_4$, $a_5$ in linear order. The problem instance admits the stable matching $M=\{\{a_1, a_2\}, \{a_1, a_3\}, \{a_2, a_3\}, \{a_4, a_5\}\}$. Note that if all agents had capacity 1 instead, i.e., if the instance were equivalent to an {\sc sr} instance, then it would be unsolvable. On the other hand, Table \ref{table:unsolvablesf} shows an unsolvable {\sc sf} instance (with a generalised stable partition indicated; a structure which we introduce later in this paper).

\begin{table}[!htb]
\centering
\centering
\caption{A solvable {\sc sf} instance with a stable matching (or an equivalent stable partition consisting of transpositions only) indicated using boxes.}
    \begin{tabular}{ c | c | c c c c }
    $a_i$ & $c_i$ & pref \\\hline
    $a_1$ & 2 & $\boxed{a_2}$ & $\boxed{a_3}$ & $a_4$ & $a_5$ \\
    $a_2$ & 2 & $\boxed{a_1}$ & $\boxed{a_3}$ & $a_5$ & $a_4$ \\
    $a_3$ & 2 & $\boxed{a_1}$ & $\boxed{a_2}$ & $a_4$ & $a_5$ \\
    $a_4$ & 2 & $a_1$ & $a_2$ & $a_3$ & $\boxed{a_5}$ \\ 
    $a_5$ & 2 & $a_2$ & $a_1$ & $a_3$ & $\boxed{a_4}$ 
    \end{tabular}
\label{table:solvablesf}
\end{table}

\begin{table}[!htb]
\centering
\caption{An unsolvable {\sc sf} instance with a generalised stable partition indicated using boxes (for transpositions), solid circles (for non-transposition successors), and broken circles (for non-transposition predecessors).}
    \begin{tabular}{ c | c | c c c c }
    $a_i$ & $c_i$ & pref \\\hline
    $a_1$ & 2 & \circled{$a_2$} & $\boxed{a_4}$ & \dottedcircled{$a_3$} & $a_5$ \\
    $a_2$ & 2 & $\boxed{a_4}$ & \circled{$a_3$} & \dottedcircled{$a_1$} & $a_5$ \\
    $a_3$ & 2 & \circled{$a_1$} & $\boxed{a_5}$ & \dottedcircled{$a_2$} & $a_4$ \\
    $a_4$ & 2 & $a_3$ & $\boxed{a_1}$ & $\boxed{a_2}$ & $a_5$ \\
    $a_5$ & 1 & $\boxed{a_3}$ & $a_1$ & $a_2$ & $a_4$ 
    \end{tabular}
\label{table:unsolvablesf}
\end{table}
\end{example}

A recent work by \citet{manipulation24} expanded on instance modification techniques that lead to stable outcomes. Specifically, for an {\sc sr} instance $I=(A,\succ)$, the authors define the concept of a \emph{removable set} of agents $S$, i.e., a set of agents $S\subseteq A$ such that $I'=(A\setminus S, \succ')$, where $\succ'$ is restricted to agents in $A\setminus S$, is solvable. They prove that given a subset of agents $T\subseteq A$, it is NP-complete to decide whether there exists a removable set $S\subseteq T$, which also gives a negative answer to a question of \citet{tan91_1} on the possibility of characterizing smallest removable sets with fixed agents (in contrast to allowing arbitrary agents to be removed which is solvable in polynomial time). Furthermore, in the context of {\sc sf}, the authors state that if the aim is to decrease the vertex capacities by a minimum total amount, then the problem is polynomial-time solvable by reducing the instance to {\sc sr}. However, the authors prove that, in general, it is NP-complete to decide whether there exists a removable set of agents of size at most $k$, even if all capacities are at most 3. This is interesting because it follows that even in this restricted {\sc sf} setting, finding a minimum-sized set of agents to delete from the instance to leave a solvable sub-instance is NP-hard, in contrast to the case for {\sc sr}.

The study of near-feasible stable matchings was initiated by Nguyen and Vohra \citep{nguyen18} (for problem models other than {\sc sf}), and has since received significant attention (e.g., by Csáji et al. \citep{couples24}). For {\sc sf}, Csáji \citep{gergely25} recently provided a polynomial-time algorithm to find a near-feasible stable matching where each agent's capacity is modified by at most 1 and such that the sum of all (positive and negative) capacity changes is at most 1.

There are also other generalisations of {\sc sr} that have been studied in the literature. For example, various works by Biró and Fleiner \citep{BiroPhD08,biro10,fleinerphd18} considered the {\sc Integral Stable Allocation} problem ({\sc isa}), which is a generalisation of {\sc sf} in which both agents and edges have upper capacity constraints. \citet{deanmunshi10} studied the real-valued generalisation of this problem, referred to as the {\sc Stable Allocation} problem ({\sc sa}). Cechlárová and Borbel’ová \citep{bmatchingrotations,sma05} and Cechlárová and Fleiner \citep{cechlarova05} studied the {\sc Stable Multiple Activities} problem ({\sc sma}), another generalisation of {\sc sf} which is a special case of {\sc isa} in which all edges have capacity 1, but parallel edges between agents are allowed to account for different activities.

\subsection{Stable Half-Matchings}

Another problem related to {\sc sf} that has been studied is the stable half-matching problem. Fleiner \citep{Fleiner02,Fleiner08} seems to have been the first to study stable fractional matchings in the many-to-many non-bipartite matching model with linear preference orders. The author assumes the following model: a \emph{graphic preference system} $(G,\succ, c)$ where $G$ is a graph consisting of vertices $V(G)$ and edges $E(G)$ and, for every agent $a_i\in V(G)$, $\succ_i$ is a linear preference order over all edges $E(a_i)\subseteq E(G)$ incident to $a_i$ in $G$. Furthermore, the author assumes a capacity function $c : V(G) \rightarrow \mathbb{N}$ (the author uses $b$ rather than $c$). Note that if we do not allow parallel edges and assume a complete graph, then a graphic preference system with \emph{strict} preference orders is equivalent to the preference list model assumed in {\sc sf}. A function $w : E(G) \rightarrow \mathbb{R}^{\geq 0}$ is called a \emph{fractional matching} if $\sum_{e\in E(G) : a_i\in e} w(e)\leq c(a_i)$ for all $a_i\in V(G)$. A fractional matching is a (\emph{half-integral}) matching if $w(e)\in \{0,1\}$ (respectively $w(e)\in \{0,\frac{1}{2}, 1\}$) for all $e\in E(G)$. Furthermore, $w$ is called \emph{stable} in \citep{Fleiner02} if for every edge $e\in E(G)$, either $w(e)=1$ or $e$ contains a vertex $a_i\in V(G)$ such that $\sum_{f\in E(G): a_i\in f ; f\succeq_i e} w(f) = c(a_i)$. From here on, we will refer to a half-integral fractional matching as a half-matching and will also use the notation $M^{\text{half}}$ and $M^{\text{full}}$ to represent the sets of unordered agent pairs that are half-matched (i.e., $M^{\text{half}}=\{e\in E(G)\;\vert\;w(e)=\tfrac{1}{2}\}$) and fully matched (i.e., $M^{\text{full}}=\{e\in E(G)\;\vert\;w(e)=1\}$), respectively, in a half-integral weight function $w$.\footnote{Note that if edges $e=\{v_i,v_j\}\in E(G)$ were allowed to have (non-unit) non-negative real-valued capacities $c(e)\in [0,\min\{c(v_i),c(v_j)\}]$, and $w(e)$ were not required to be in $[0,1]$ but rather in $[0,c(e)]$, then we would recover (a parallel-edge generalisation of) the {\sc sa} problem \citep{deanmunshi10}.} Fleiner gave the following characterisation of stable half-matchings.

\begin{theorem}[\citep{Fleiner08}]
\label{thm:fleinerstructure}
    In a graphic preference system $(G,\succ, c)$, a stable half-matching always exists and consists of disjoint subsets $M^{\text{full}}$, $M^{\text{half}}$ of edges such that
    \begin{itemize}
        \item the components of $M^{\text{half}}$ are cycles;
        \item for any vertex $a_i\in V(G)$, $\vert E(a_i)\cap M^{\text{full}} \vert + \frac{1}{2} \vert E(a_i) \cap M^{\text{half}}\vert\leq c(a_i)$;
        \item for any $a_i\in V(G)$, if $a_i$ is incident to some edge in $M^{\text{half}}$, then $\vert E(a_i)\cap M^{\text{full}} \vert + \frac{1}{2} \vert E(a_i) \cap M^{\text{half}}\vert = c(a_i)$; 
        \item each edge $e\in E(G)\setminus M^{\text{full}}$ has at least one vertex $a_i$ where $\vert E(a_i)\cap M^{\text{full}} \vert + \frac{1}{2} \vert E(a_i) \cap M^{\text{half}}\vert = c(a_i)$ and $a_i$ strictly prefers all edges $E(a_i)\cap (M^{\text{full}}\cup M^{\text{half}})$ to $e$.
    \end{itemize}
\end{theorem}

Theorem \ref{thm:fleinerstructure}, which \citet{Fleiner08} proved using Scarf's Lemma, already suggests that stable partitions from {\sc sr} might generalise to {\sc sf}. It is important to note that Scarf's Lemma is very powerful and has already been used to show that a stable half-matching always exists for a range 
of stable matching problems, including {\sc sf}, for example, by \citet{aharonifleiner03}, and \citet{biro16}. \citet{Fleiner08} also proved the following result using a transformation from {\sc sf} to {\sc sr}.

\begin{theorem}[\citep{Fleiner08}]
\label{thm:fleinerinvariance}
    In a graphic preference system $(G,\succ, c)$, if $w$ is a stable half-matching of $(G,\succ, c)$ and there exists a cycle of edges $e_{i_1}, e_{i_2}, \dots, e_{i_k}$ in $E(G)$ such that $w(e_{i_s})=\frac{1}{2}$ for all $1\leq s\leq k$ and $k$ is odd, then for any stable half-matching $w'$ of $(G,\succ, c)$, it is the case that $w'(e_{i_s})=\frac{1}{2}$.
\end{theorem}

Although the results above focus on existence and connections to fixed-point theorems, they provide hope for a generalisation of stable partitions in the many-to-many setting. A characterisation of {\sc sf} instances through stable partitions in the spirit of Tan, together with an exploration of their structure and suitability as a solution concept for {\sc sf}, has, to the best of our knowledge, not yet been pursued. The closest is work by \citet{Fleiner2010}, who extended the above-mentioned study of stable half-matchings in a many-to-many non-bipartite setting to a more general model where preferences are expressed through \emph{choice functions} (which need not induce linear preference orders in general). We will return to this work briefly in Sections \ref{sec:GSPalgorithm} and \ref{sec:oddcycles}. Still, we believe it is worthwhile to investigate how stable partitions extend from {\sc sr} to the many-to-many non-bipartite matching model and formalise them independently to reason about their structure in the hope of finding new structural and algorithmic results. For example, can we characterise what we see in Table \ref{table:unsolvablesf} as some form of stable partition $\Pi=(a_1 \; a_2 \; a_3)(a_1 \; a_4)(a_2 \; a_4)(a_3 \; a_5)$ and, if so, can it be computed efficiently? We provide answers to these questions, and others, in the following sections.

\section{Towards Generalised Stable Partitions}
\label{sec:definingGSP}

With stable partitions of {\sc sr} instances being permutations of the agents, there is no unique trivial extension of Definition \ref{def:sp} to the concept of a \emph{generalised stable partition} (GSP) in the many-to-many setting. We have some basic requirements: a GSP should extend the concept of a stable partition for {\sc sr} instances to {\sc sf} instances, satisfy similar properties, and correspond to a half-matching with some sort of stability condition. We will start by proposing two different definitions for a GSP, denoted by GSP1 and GSP2, show some interesting structural properties, and end this section by highlighting their equivalence.

\subsection{Definition of GSP1 in Terms of Nested Permutations}

For the following definition of a GSP1, a \emph{cyclic permutation} (also referred to as a \emph{$k$-perm} for a cyclic permutation with cycle length $k$) is a permutation consisting of a single cycle.\footnote{Note that some authors consider a permutation to be cyclic when written in cyclic notation, which is not sufficient here for our purposes. For example, for a set $S=\{x,y,z\}$, we would consider the permutations $(x\;y\;z)$ and $(x\;z\;y)$ to be cyclic permutations of $S$ because they consist of a single cycle. On the other hand, the permutations $(x\; y)(z)$ and $(x)(y)(z)$ consist of more than one cycle, so we do not consider them to be cyclic permutations of $S$. The GSP1 structure is defined in terms of multiple cyclic permutations on subsets, so this assumption does not come at a loss of generality: we can simply construct $(x\; y)(z)$ through two cyclic permutations $\Pi_1=(x\;y)$ and $\Pi_2=(z)$ on subsets $S_1=\{x,y\}$ and $S_2=\{z\}$, respectively.} Furthermore, we consider two permutations to be \emph{distinct} when there exists an element that is mapped to two different elements in the respective permutations.

\begin{definition}[GSP1]
\label{def:gsp1}
    Let $I=(A,\succ, c)$ be an {\sc sf} instance. Then a \emph{GSP1} $\Pi$ is a collection of $k$ distinct (apart from fixed points) cyclic permutations $\Pi_i$ (for $1\leq i\leq k$) of sets $A_i\subseteq A$ such that  
    \begin{enumerate}[leftmargin=3.5em]
        \item[(F1)] $\forall \; \Pi_i\in \Pi$ and $\forall \; a_j\in A_i$ we have $\Pi_i(a_j)\succeq_j \Pi_i^{-1}(a_j)$, 
        \item[(F2)] $\nexists \; a_i, a_j \in A$ where $a_i\neq a_j$ and $(a_i\; a_j)\notin \Pi$ such that $a_j \succ_i \Pi_r^{-1}(a_i)$ and $a_i \succ_j \Pi_s^{-1}(a_j)$ for some $\Pi_r, \Pi_s\in \Pi$,
        \item[(F3)] $\forall a_i\in A$ we have $\vert \{r \; \vert \; a_i\in A_r\}\vert = c_i$, and
        \item[(F4)] $\forall a_i, a_j \in A$ with $a_i\neq a_j$ we have that $\vert \{s\in \{1,2,\dots,k\}\; \vert \; \Pi_s(a_i) = a_j\} \vert + \vert \{s \; \vert \; \Pi_s(a_j) = a_i\} \vert \leq 2$.
    \end{enumerate}
    We call a GSP1 \emph{reduced} if it does not contain any cycles of even length longer than 2.
\end{definition}

A GSP1 can be thought of as a \emph{permutation with layers} (rather than merely a composition of cycles). For example, the instance shown in Table \ref{table:unsolvablesf} admits the GSP1 $\Pi=(a_1 \; a_2 \; a_3)_1(a_1 \; a_4)_2(a_2 \; a_4)_3(a_3 \; a_5)_4$ consisting of permutations $\Pi_1=(a_1 \; a_2 \; a_3)_1$, $\Pi_2=(a_1 \; a_4)_2$, $\Pi_3=(a_2 \; a_4)_3$ and $\Pi_4=(a_3 \; a_5)_4$. By abuse of notation, we will sometimes use $\Pi=\{\Pi_1,\Pi_2,\Pi_3,\Pi_4\}$ and $\Pi=\Pi_1\Pi_2\Pi_3\Pi_4$ interchangeably.

Conditions F1 (``every agent weakly prefers their successor to their predecessor in every cycle'') and F2 (``no two distinct agents that are not in some transposition strictly prefer each other over some other agents they are assigned to'') are direct stability analogues of T1 and T2 in the case of {\sc sr}, whereas F3 is necessary to ensure that no agent exceeds their capacity when nesting the permutations. Furthermore, F4 ensures that no two agents are assigned more than value one (two successor-predecessor relations) to each other. By inspection, none of the conditions is redundant. We call two distinct agents $a_i,a_j$ \emph{blocking} if they violate F2. 

We consider a half-matching $M$ corresponding to a GSP1 $\Pi$ by letting any predecessor-successor pair in each cycle of $\Pi$ be a \emph{half-match}. Two half-matches form a \emph{full match}. Note that in contrast to the {\sc sr} setting, an agent $a_i\in A$ could be fully matched to some agent $a_j$ in a collection of permutations $\Pi=\{\Pi_k\}$ if either $\Pi_r(a_i)=\Pi_s^{-1}(a_i)=a_j$, $\Pi_r^{-1}(a_i)=\Pi_s^{-1}(a_i)=a_j$ for $\Pi_r\neq\Pi_s$, or $\Pi_r(a_i)=\Pi_s(a_i)=a_j$ (for $\Pi_r\neq\Pi_s$). However, not all of these cases are admissible for a GSP1. For example, we want to rule out that odd cycles can be interchanged with even cycles: $\Pi = (a_1 \; a_2 \; a_3)(a_2 \; a_1 \; a_4)$, as a half-matching, is equivalent to $\Pi' = (a_1 \; a_2)(a_2 \; a_3 \; a_1 \; a_4)$, where both $\Pi, \Pi'$ can simultaneously satisfy F1 and F3 (by inspection). However, the following results show that this cannot happen due to F2 and F4.

\begin{lemma}
\label{lemma:equal1}
    Let $\Pi$ be a GSP1 containing $\Pi_r,\Pi_s$. Then for all $a_i, a_j\in A$ where $a_i\neq a_j$, if $\Pi_r^{-1}(a_j)=a_i$ and $\Pi_s^{-1}(a_j)=a_i$, we must have $\Pi_r = \Pi_s$.
\end{lemma}
\begin{proof}
    Suppose that $\Pi_r\neq \Pi_s$ and let $a_k=\Pi_r^{-1}(a_i)$ and let $a_k'=\Pi_s^{-1}(a_i)$. 
    
    By F4, clearly $a_j\notin\{a_k, a_k'\}$. For now, suppose that $a_k\neq a_k'$. Without loss of generality, suppose furthermore that $a_k\succ_i a_k'=\Pi_s^{-1}(a_i)$. By F1, we know that $a_i=\Pi_r(a_k)\succ_k \Pi_r^{-1}(a_k)$. Finally, we also know that $(a_i\; a_k)\notin\Pi$, otherwise F4 would be violated. Therefore, $a_i,a_k$ violate F2, a contradiction of stability of $\Pi$, thus we must have $\Pi_r=\Pi_s$ as desired.

    Consider the case where $a_k=a_k'$. Now, we can look at the predecessors of $a_k,a_k'$ in $\Pi_s,\Pi_r$, and either they are distinct, in which case we can apply the same argument as above, or they are also equal, in which case we can take another step back. Now notice that if $\Pi_r\neq \Pi_s$ then by definition of a stable partition, $\Pi_r$ and $\Pi_s$ must be distinct permutations consisting of one cycle, so this sequence of equal predecessors cannot continue indefinitely and the result follows.
\end{proof}

Furthermore, we can also show that no two agents can be flipped predecessor-successor pairs in different cycles of the same GSP1.

\begin{lemma}
\label{lemma:equal2}
    For all $a_i, a_j\in A$ where $a_i\neq a_j$, if $\Pi_r(a_j) = a_i$ and $\Pi_s(a_i)=a_j$ for GSP1 $\Pi$ containing $\Pi_r$ and $\Pi_s$, then $\Pi_r = \Pi_s$.
\end{lemma}
\begin{proof}
    Suppose that $\Pi_r\neq\Pi_s$, then by F4 we must have $\Pi_r^{-1}(a_j)\neq a_i$ and $\Pi_s^{-1}(a_i)\neq a_j$. By F1, clearly $a_i\succ_j \Pi_r^{-1}(a_j)$ and $a_j\succ_i \Pi_s^{-1}(a_i)$. However, by F4 we must also have that $(a_i\; a_j)\notin\Pi$. Thus, $a_i,a_j$ violate F2, a contradiction of stability of $\Pi$, so we must have $\Pi_r=\Pi_s$ as desired.
\end{proof}

Therefore, it follows immediately that any two agents that are fully matched in a GSP1 cannot be half-matched in different cycles.

\begin{theorem}
\label{thm:fullymatched}
     Let $a_i \neq a_j\in A$ be fully matched in the half-matching associated with GSP1 $\Pi$. Then there exists a permutation $\Pi_r\in \Pi$ such that $\Pi_r=(a_i\; a_j)$.
\end{theorem}
\begin{proof}
    By $a_i, a_j$ matched, we know that there exists some $\Pi_r\in \Pi$ such that either $\Pi_r^{-1}(a_j)=a_i$ or $\Pi_r^{-1}(a_i)=a_j$. Suppose (wlog) that $\Pi_r^{-1}(a_j)=a_i$. By being fully matched, either $\Pi_t^{-1}(a_j)=a_i$ for some $\Pi_t\neq \Pi_r$ or $\Pi_t(a_j)=a_i$ for any $\Pi_t\in\Pi$. The former case cannot happen due to Lemma \ref{lemma:equal1}. Thus, $\Pi_t(a_j)=a_i$, but we know that $\Pi_r(a_i)=a_j$ by assumption, so $\Pi_r=\Pi_t$ by Lemma \ref{lemma:equal2}. Uniqueness of $\Pi_r$ follows from F4.
\end{proof}

\begin{corollary}
\label{cor:notwoconsecutive}
     In a GSP1 $\Pi$, no two distinct agents $a_i,a_j$ can be present consecutively in two separate cycles.
\end{corollary}
\begin{proof}
    Suppose that they appear consecutively in two distinct cycles $C_1,C_2$. Clearly, neither cycle can be a transposition because F4 would be violated, and by the same argument, there cannot exist a distinct third cycle in which $a_i,a_j$ can be present consecutively. Also, $C_1$ and $C_2$ each contribute a 0.5-weighted match $\{a_i,a_j\}$ to the half-matching $M$ associated with $\Pi$. However, then $\{a_i,a_j\}$ must be a full match, so by Theorem \ref{thm:fullymatched}, it must be that $(a_i\;a_j)\in\Pi$, a contradiction.
\end{proof}

Furthermore, we can use a similar argument to greatly restrict the number of long cycles.

\begin{theorem}
\label{thm:atmostonelonger}
    Any agent $a_i\in A$ can be in at most one cyclic permutation of length longer than 2 in any GSP1 $\Pi$. 
\end{theorem}
\begin{proof}
    Suppose there exists an agent $a_i$ who is contained in at least two such cyclic permutations $\Pi_r, \Pi_s$ in $\Pi$. By definition, $\Pi_r$ and $\Pi_s$ are distinct and by Theorem \ref{thm:fullymatched}, for every agent $a_i\in A(\Pi_r)\cup A(\Pi_s)$, we must have $\{\Pi_r^{-1}(a_i),\Pi_r(a_i)\}\cap \{\Pi_s^{-1}(a_i),\Pi_s(a_i)\}=\varnothing$. Suppose (wlog) that $a_i$ prefers their predecessor in $\Pi_r$ to their predecessor in $\Pi_s$ and denote the predecessors by $a_r,a_s$, respectively, i.e., $a_r\succ_i a_s=\Pi_s^{-1}(a_i)$. By F1 and the length of $\Pi_r$ being above 2, we know that $a_i\succ_r \Pi_r^{-1}(a_r)$. Thus, $a_i$ and $a_r$ violate F2, a contradiction, so the result follows.
\end{proof}

Because a GSP1 generalises the concept of a stable partition, we observe the following.

\begin{lemma}
\label{lemma:spgspcorrespondence}
    Let $I=(A,\succ)$ be an {\sc sr} instance and define $J=(A, \succ, c)$ as the corresponding {\sc sf} instance in which $c_i=1$ for all $a_i\in A$. Then $\Pi$ is a stable partition of $I$ if and only if $\{\Pi\}$ is a GSP1 of $J$.
\end{lemma}
\begin{proof}
    It is easy to see that in this case, T1 holds if and only if F1 holds, and that F3 and F4 are trivially satisfied. Also, as there is only one permutation $\Pi$, T2 holds if and only if F2 holds.
\end{proof}

\subsection{Definition of GSP2 in Terms of Set Functions}

The following is a different approach to formalising a GSP in terms of set functions $S,P$ (for \emph{s}uccessor and \emph{p}redecessor), rather than permutations. First, we define the following notation: let $a_i$ be an agent that has a strict preference relation $\succ_i$ over a set of agents $S=\{a_1, a_2, \dots,a_j\}$, then worst$_i(S)$ is the worst-ranked agent in $S$ (in the preference relation of $a_i$), i.e., agent $a_r\in S$ such that for all $a_s\in S$ we have $a_s\succeq_i a_r$.

\begin{definition}[GSP2]
    Let $I=(A,\succ, c)$ be an {\sc sf} instance. Then a \emph{GSP2} $\Pi=(S,P)$ consists of functions $S,P:A\rightarrow \mathcal{P}(A)$ such that   
    \begin{enumerate}[leftmargin=3.5em]
        \item[(G1)] $\forall \; a_i\in A. \; \exists$ bijection $f : P(a_i)\rightarrow S(a_i)$ such that $\forall a_j\in P(a_i)$ we have $f(a_j)\succeq_i a_j$, 
        \item[(G2)] $\nexists \; a_i, a_j \in A$ with $a_i\neq a_j$ and $a_i\notin S(a_j)\cap P(a_j)$ such that $a_i\succ_j$ worst$_j(P(a_j))$ and $a_j\succ_i$ worst$_i(P(a_i))$,  
        \item[(G3)] $\forall \; a_i \in A$ we have $\vert P(a_i)\vert=\vert S(a_i)\vert = c_i$, and 
        \item[(G4)] $\forall \; a_i, a_j \in A$ we have $a_i\in P(a_j)$ if and only if $a_j\in S(a_i)$.
    \end{enumerate}
\end{definition}

Here, the correspondence with half-matchings comes even more naturally. We let two agents $a_i, a_j$ be \emph{half-matches} if either $a_i\in S(a_j)$ or $a_i\in P(a_j)$ (but not both) and be \emph{full matches} if $a_i\in S(a_j)\cap P(a_j)$ (or equivalently for $a_j$ in terms of $a_i$'s sets due to G4). We call two distinct agents $a_i,a_j$ \emph{blocking} with regards to $(S,P)$ if they violate G2.

\begin{example}
To illustrate a GSP2, consider the functions $S,P$ given in Table \ref{table:gsp2example} for the instance shown in Table \ref{table:unsolvablesf}. The structure corresponds to the GSP1 example above and satisfies all conditions G1-G4, as can be verified. For example, G3 and G4 can be verified by inspection of the sets and, for G1, taking agent $a_1$ as an example, there exists bijection $f$ from $P(a_1)$ to $S(a_1)$ given by $f(a_3)=a_4$, $f(a_4)=a_2$ (notice that the GSP1 previously given for the instance would correspond to $f(a_3)=a_2$, $f(a_4)=a_4$ which also satisfies our conditions). Lastly, G4 can be verified by inspecting the preference lists.

\begin{table}[!htb]
\centering
\caption{A GSP2 for the Instance from Table \ref{table:unsolvablesf}; we can consider this a function from $a_i$ to its predecessor and successor sets $P(a_i),S(a_i)$.}
\begin{tabular}{ c | c | c  }
$a_i$ & $P(a_i)$ & $S(a_i)$ \\\hline
$a_1$ & $\{ a_3, a_4 \}$ & $\{ a_2, a_4 \}$ \\
$a_2$ & $\{ a_1, a_4 \}$ & $\{ a_3, a_4 \}$ \\
$a_3$ & $\{ a_2, a_5 \}$ & $\{ a_1, a_5 \}$ \\
$a_4$ & $\{ a_1, a_2 \}$ & $\{ a_1, a_2 \}$ \\
$a_5$ & $\{ a_3 \}$ & $\{ a_3 \}$
\end{tabular}
\label{table:gsp2example}
\end{table}
\end{example}

Note that in this setting, we need to be careful about \emph{fixed-points}, i.e., agents that are assigned to themselves. For example, let $I$ be an {\sc sf} instance with two agents $a_1, a_2$, each with capacity 3 (here, we allow $c_i>n$ for illustrative purposes). Then, we would expect that the agents are fully matched, but have two values of free capacity each. In the {\sc sr} context (due to the completeness of the preference lists), an agent can have at most one value of free capacity in a stable half-matching, expressed through a cycle of length 1 of the agent in the associated stable partition due to the assumption that every agent ranks themselves last (i.e., every agent would rather be matched to any other agent in their preference list rather than being unmatched). In a GSP1, we can simply have the same agent in multiple cycles of length 1 without violating condition F4 because $a_i=a_j$. However, this will not work in a GSP2 because there is no natural way to deal with multi-unit self-allocations in terms of sets. We can deal with this as follows: for every agent $a_i$ with capacity $c_i$, we will introduce $c_i$ dummy agents $d_i^{j}$ for $1 \leq j \leq c_i$, each with capacity 1. We will add all $d_i^{j}$ agents to the end of $a_i$'s preference list in order of $j$. Each $d_i^{j}$ ranks $a_i$ first and themselves second (and if we want to enforce complete preference lists, then all other agents in some arbitrary order). This way, if $a_i$ has $f_i\leq c_i$ free allocation values, then $a_i$ is in $f_i$ full matches with copies of itself, and all $c_i-f_i$ dummy agent copies of $a_i$ are fixed points. 

\begin{example}
Table \ref{table:4free} shows an {\sc sf} instance with a stable matching in which $a_1$ has all of its $c_1=4$ capacity unallocated. Table \ref{table:nofree} shows how the preference lists of this instance can be extended (including making the agent copies $d_1^{1}$ to $d_1^{4}$) such that there exists a complete stable matching (or a stable partition) with no free values. In practice, we will assume that the dummy agents were added where necessary.

\begin{table}[!htb]
\centering
\centering
\caption{A stable matching where $a_1$ has free capacity 4.}
    \begin{tabular}{ c | c | c c c c }
    $a_i$ & $c_i$ & pref \\\hline
    $a_1$ & 4 & $a_2$ & $a_3$ & $a_4$ & $a_5$ \\
    $a_2$ & 1 & $\boxed{a_3}$ & $a_1$ & $a_4$ & $a_5$ \\
    $a_3$ & 1 & $\boxed{a_2}$ & $a_1$ & $a_4$ & $a_5$ \\
    $a_4$ & 1 & $\boxed{a_5}$ & $a_1$ & $a_2$ & $a_3$ \\ 
    $a_5$ & 1 & $\boxed{a_4}$ & $a_1$ & $a_2$ & $a_3$ 
    \end{tabular}
\label{table:4free}
\end{table}

\begin{table}[!htb]
\centering
\caption{A stable matching in a modified instance where no agent has any free capacity.}
    \begin{tabular}{ c | c | c c c c c c c }
    $a_i$ & $c_i$ & pref \\\hline
    $a_1$ & 4 & $a_2$ & $\dots$ & $a_5$ & $\boxed{d_1^{1}}$ & $\boxed{d_1^{2}}$ & $\boxed{d_1^{3}}$ & $\boxed{d_1^{4}}$ \\
    $d_1^{1}$ & 1 & $\boxed{a_1}$ & $d_1^{1}$ & $\dots$ \\
    $d_1^{2}$ & 1 & $\boxed{a_1}$ & $d_1^{2}$ & $\dots$ \\
    $d_1^{3}$ & 1 & $\boxed{a_1}$ & $d_1^{3}$ & $\dots$ \\
    $d_1^{4}$ & 1 & $\boxed{a_1}$ & $d_1^{4}$ & $\dots$ \\
    $a_2$ & 1 & $\boxed{a_3}$ & $a_1$ & $a_4$ & $a_5$ \\
    $a_3$ & 1 & $\boxed{a_2}$ & $a_1$ & $a_4$ & $a_5$ \\
    $a_4$ & 1 & $\boxed{a_5}$ & $a_1$ & $a_2$ & $a_3$ \\ 
    $a_5$ & 1 & $\boxed{a_4}$ & $a_1$ & $a_2$ & $a_3$
    \end{tabular}
\label{table:nofree}
\end{table}
\end{example}

\begin{example}
An interesting behaviour that does not occur in the {\sc sr} model is shown in Table \ref{table:curious}. If $c_1$ was 1, then we would have an {\sc sr} instance admitting stable partitions (GSP1s) $\Pi_1 = (a_1 \; a_2)(a_3 \; a_4)$, $\Pi_2 = (a_1 \; a_4)(a_2 \; a_3)$, and $\Pi_3 = (a_1 \; a_2 \; a_3 \; a_4)$. However, now that $a_1$ has one additional value of capacity, while $\Pi_2\cup\{(a_1)\}$ is stable (i.e., agent $a_1$ has free capacity 1), both $\Pi_1$ and $\Pi_3$ cannot accommodate this change in a similar way (free capacity of one for $a_1$ would mean that $a_1$ blocks with $a_4$). The fact that $\Pi_3$ is not preserved as a stable partition when increasing $a_1$'s capacity is reflected by the following statement.

\begin{table}[!htb]
\centering
\caption{An {\sc sf} instance with a unique GSP, while $c_1=1$ gives three distinct stable partitions.}
    \begin{tabular}{ c | c | c c c }
    $a_i$ & $c_i$ & pref \\\hline
    $a_1$ & 2 & $a_2$ & $\boxed{a_4}$ & $a_3$ \\
    $a_2$ & 1 & $\boxed{a_3}$ & $a_1$ & $a_4$ \\
    $a_3$ & 1 & $a_4$ & $\boxed{a_2}$ & $a_1$ \\
    $a_4$ & 1 & $\boxed{a_1}$ & $a_3$ & $a_2$ 
    \end{tabular}
\label{table:curious}
\end{table}
\end{example}

The following lemma makes an interesting observation regarding dummy agents.
 
\begin{lemma}
\label{lemma:matchedtocapacity}
    If some agent $a_i$ is half-matched to agents $a_j,a_k$ in the half-matching of a GSP2 $(S,P)$, then $\vert S(a_i)\setminus D(a_i)\vert = \vert P(a_i)\setminus D(a_i)\vert = c_i$, where $D(a_i)$ is the set of dummy agents for $a_i$, i.e., $a_i$ is matched up to capacity without dummy agents.
\end{lemma}
\begin{proof}
    Suppose that $\vert P(a_i)\setminus D(a_i)\vert < c_i$. By construction, we know that all dummy agents $D(a_i)$ are ranked strictly worse by $a_i$ than any non-dummy agents, i.e., $a_k\succ_i d_i^{l}$ for all $1\leq l \leq c_i$. Therefore, worst$_i(P(a_i))= d_i^{l}$ for some $1\leq l \leq c_i$.

    Suppose that one of $a_j$ or $a_k$ is in $P(a_i)$. Without loss of generality, take $a_j\in P(a_i)$, then by G4 we must have $a_i\in S(a_j)$. By being half-matched to $a_i$, we also know that $a_i\notin P(a_j)$, so by G1 we must have $a_l\in P(a_j)$ such that $a_i\succ_j a_l \succeq_j$ worst$_j(P(a_j))$. Therefore $a_i,a_j$ violate G2, a contradiction, so we must have $\{a_j,a_k\}\subseteq S(a_i)$.

    By $a_j,a_k$ being half-matched to $a_i$ and $\{a_j,a_k\}\subseteq S(a_i)$, we must have that $\{a_j', a_k'\}\subseteq P(a_i)\setminus S(a_i)$ by G1 such that $a_j'\neq a_k'$, $a_j\succ_i a_j'$ and $a_k\succ_i a_k'$. Suppose (wlog) that $a_j'\succ_i a_k'\succeq_i$ worst$_i(P(a_i))$. By G4 and the construction, we must have that $a_i\in S(a_j')\setminus P(a_j')$, so by G1 we must have $a_i\succ_{j'}$ worst$_{j'}(P(a_j'))$. Thus, $a_i,a_j'$ violate G2, a contradiction, so we must have $\vert P(a_i)\setminus D(a_i)\vert \geq c_i$. 

    Furthermore, the dummy agents are always ranked last, so together with G1, we must have that $\vert S(a_i)\setminus D(a_i)\vert\geq \vert P(a_i)\setminus D(a_i)\vert$, but due to G3 clearly $c_i \geq \vert S(a_i)\setminus D(a_i)\vert$, so indeed $\vert S(a_i)\setminus D(a_i)\vert = \vert P(a_i)\setminus D(a_i)\vert = c_i$ as desired.
\end{proof}

\subsection{Equivalence of GSP1 and GSP2}

We now show that the two definitions of GSPs presented above are equivalent and can thus be used interchangeably. We start with one direction.

\begin{lemma}
    Let $I=(A,\succ, c)$ be an {\sc sf} instance, let $\Pi=\{\Pi_1,\dots,\Pi_k\}$ be a GSP1 of $I$, and construct $(S,P)$ in the following way: for every $\Pi_r\in \Pi$ and every $a_i\in A_r$, let $a_j=\Pi_r(a_i)$ (if $a_i = a_j$, replace $a_j$ by a dummy agent of $a_i$), add $a_j$ to $S(a_i)$ and add $a_i$ to $P(a_j)$. Then, upon termination of this process, $(S,P)$ is a GSP2.
\end{lemma}
\begin{proof}
    G1 is satisfied by construction, because for every $a_j\in A$, if $a_i\in P(a_j)$ then $a_i=\Pi_r^{-1}(a_j)$ for some $\Pi_r\in \Pi$. Thus, there must exist some corresponding $a_k=\Pi_r(a_j)\in S(a_j)$ such that $a_k\succeq_j a_i$ by F1.

    Suppose that G2 does not hold. Then there exist agents $a_i\neq a_j$ not fully matched to each other such that $a_j$ prefers $a_i$ to their worst predecessor $a_w$ in $P(a_j)$ and $a_i$ prefers $a_j$ to their worst predecessor $a_w'$ in $P(a_i)$. However, by construction, $a_i$ and $a_j$ are also not fully matched in $\Pi$ and $a_w=\Pi_r^{-1}(a_j)$, $a_w'=\Pi_s^{-1}(a_i)$ for some $\Pi_r, \Pi_s \in \Pi$. As this cannot happen due to F2, G2 must hold.
    
    Next, we have that, by construction, for any agent $a_i\in A$, $P(a_i) = \{a_j^*\in A \; \vert \; \exists\; \Pi_r\in \Pi \text{ such that } \Pi_r^{-1}(a_i)=a_j\}$ and $S(a_i) = \{a_j^*\in A \; \vert \; \exists\; \Pi_r\in \Pi \text{ such that } \Pi_r(a_i)=a_j\}$ (where $a_j^*$ indicates that the agent could be replaced by a dummy agent in the construction if $a_i=a_j$). Also, $\vert\{a_j^* \in A\; \vert \; \exists\; \Pi_r\in \Pi \text{ such that }\Pi_r^{-1}(a_i)=a_j\}\vert = \vert\{1\leq r\leq k \; \vert \; a_i \in A_r\}\vert = \vert\{a_j^* \; \vert \; \exists\;\Pi_r\in \Pi \text{ such that }\Pi_r(a_i)=a_j\}\vert$ (where $k$ is the number of cyclic permutations in $\Pi$), so by F3, we must have $\vert P(a_i)\vert = c_i = \vert S(a_i)\vert$, establishing G3. 
    
    Finally, G4 holds by construction of $S$ and $P$.
\end{proof}

To show the other direction of the correspondence, it will be useful to establish the analogue of Theorem \ref{thm:atmostonelonger}, i.e.,  every agent's predecessor and successor sets differ by at most 1, as the following result shows.

\begin{lemma}
\label{lemma:intersectionuniongsp2}
    If $(S,P)$ is a GSP2 then for any agent $a_i$, $\vert S(a_i)\cap P(a_i)\vert \geq \vert S(a_i)\cup P(a_i)\vert - 1$. 
\end{lemma}
\begin{proof}
    Suppose that there exists an agent $a_i\in A$ such that $\vert S(a_i)\cap P(a_i)\vert < \vert S(a_i)\cup P(a_i)\vert - 1$ and recall that $S(a_i)$ and $P(a_i)$ are always of equal size.
    
    By G3, there exist agents $a_r\in P(a_i)$ and $a_s\in P(a_i)$ where $a_r\neq a_s$ and $a_r\notin S(a_i)$ and $a_s\notin S(a_i)$ (that is, neither of these two agents is fully matched to $a_i$). By G4, we must have $a_i\in S(a_r)\cap S(a_s)$ but $a_i\notin P(a_r)\cup P(a_s)$. 
    
    Now suppose (wlog) that $a_r\succ_i a_s$ (by strict preference relations). We know trivially that $a_s\succeq_i$ worst$_i(P(a_i))$ and thus $a_r\succ_i$ worst$_i(P(a_i))$ (by transitivity). 
    
    Finally, by G1, there is a bijection between the agents in the successor and predecessor sets of $a_r$ such that each successor is assigned to a predecessor of equal or worse rank. However, we assume that $a_i\notin P(a_r)$, therefore, there must exist a strictly worse agent $a_p\in P(a_r)$ such that $a_i\succ_r a_p$, such that $a_i\succ_r$ worst$_r P(a_r))$. Thus, agents $a_i,a_r$ contradict G2, establishing the result.
\end{proof}

\begin{corollary}
\label{corollary:differbytwo}
    For any agent $a_i$, we have $\vert S(a_i)\setminus P(a_i)\vert \in \{0,1\}$ and $\vert P(a_i)\setminus S(a_i)\vert \in \{0,1\}$, i.e., the sets $S(a_i)$ and $P(a_i)$ differ by at most two agents.
\end{corollary}

Now it is easy to see that the sets have enough structure to rebuild the permutations of a GSP1 from the GSP2 sets, which we can do using Algorithm \ref{alg:constructgsp1}. The algorithm proceeds as follows: we start with an empty GSP1 candidate $\Pi$ and extend it incrementally to form a valid GSP1. For each agent $a_i$, we first add all of its full matches (i.e., its transpositions and fixed points) as distinct permutations to $\Pi$. We refer to these partners in the full matches of $a_i$ as $a_j^*$, which could either be a dummy agent copy $d_i^k$ of $a_i$ or an agent $a_j\in A\setminus\{a_i\}$. If $a_j^*$ is indeed a dummy agent copy of $a_i$, then we assume that it is dealt with as we would expect, in the sense that we actually add the fixed-point $(a_i)$ (rather than $(a_i \; d_i^k)$).

If the agent is (also) in a cycle of length longer than 2, then by Corollary \ref{corollary:differbytwo}, there is a unique successor of $a_i$ for that cycle. But then that successor must also have a unique successor by G3, and eventually, this cycle will close back with the agent we started with. All of these intermediate agents are temporarily stored in an array $C$, and finally, this sequence of agents in $C$ is added as a cycle to $\Pi$. Recall that we assume that cycles (permutations) are equal as functions, i.e., they are equal if and only if all agents are sent to the same agents in both cycles (respectively permutations). Therefore, it does not matter whether, for example, we union our candidate $\Pi$ with $(a_{i_1} \; a_{i_2} \; a_{i_3})$, $(a_{i_2} \; a_{i_3} \; a_{i_1})$ or $(a_{i_3} \; a_{i_1} \; a_{i_2})$ (and their union is a single element), as they are equal as functions. By removing agents from the respective successor and predecessor sets when adding the cycle to $\Pi$, there is no wasted work in the sense that we do not try to add the same cycles multiple times.

\begin{algorithm}[!htb]
\renewcommand{\algorithmicrequire}{\textbf{Input:}}
\renewcommand{\algorithmicensure}{\textbf{Output:}}

    \begin{algorithmic}[1]

    \Require{$(S,P)$ : a GSP2 of an instance $I$; $A$ : the set of agents in $I$}
    \Ensure{$\Pi$ : a GSP1 of $I$}

    \State $\Pi \gets \varnothing$

    \For{agent $a_{i}\in A$}
        \For{agent $a_{j}^*\in P(a_i)\cap S(a_i)$} \Comment{Add all transpositions (and fixed-points)}
            \State $\Pi\gets\Pi\cup (a_i \; a_j^*)$
            \State $S(a_i)$.remove($a_j^*$); $P(a_i)$.remove($a_j^*$)
            \State $S(a_j^*)$.remove($a_i$); $P(a_j^*)$.remove($a_i$)
        \EndFor
    
        \If{$S(a_i)\setminus P(a_i)\neq\varnothing$} \Comment{Add the longer cycle (unique, if any)}
            \State $a_r \gets$ any$(S(a_i)\setminus P(a_i))$ \Comment{Unique by Corollary \ref{corollary:differbytwo}}
            \State $C \gets [a_i]$
            \State $S(a_i)$.remove($a_r$); $P(a_r)$.remove($a_i$)
            \While{$a_r\neq a_i$}
                \State $C$.append$(a_r)$
                \State $a_r' \gets$ any$(S(a_r)\setminus P(a_r))$ \Comment{Unique by Corollary \ref{corollary:differbytwo}}
                \State $S(a_r)$.remove($a_r'$); $P(a_r')$.remove($a_r$)
                \State $a_r\gets a_r'$
            \EndWhile
            
            $\Pi \gets \Pi\cup (C[0] \; C[1] \dots C[k-1])$, where $k=$ length$(C)$
        \EndIf
    \EndFor

    \State\Return{$\Pi$}

    \end{algorithmic}
    \caption{\texttt{ConstructGSP1}$((S,P), A)$, constructs a GSP1 $\Pi$ from a GSP2 $(S,P)$.}
    \label{alg:constructgsp1}
\end{algorithm}

Using our algorithm, we can finish the equivalence proof between the GSP concepts.

\begin{lemma}
    Let $I=(A,\succ, c)$ be an {\sc sf} instance and $(S,P)$ be a GSP2 of $I$. Then $\Pi$ computed using Algorithm \ref{alg:constructgsp1} is a GSP1 of $I$.
\end{lemma}
\begin{proof}
    We need to establish that $\Pi$ consists of distinct cyclic permutations and indeed satisfies F1-F4 in this construction. First, note that we already know by Theorem \ref{thm:atmostonelonger} that every agent $a_i$ will be in at least $c_i-1$ fixed points or transpositions in $\Pi$, all of which trivially satisfy F1. Furthermore, for the longer cycles added to $\Pi$, G1 implies F1 because the distinct successors and predecessors are unique by Lemma \ref{lemma:intersectionuniongsp2}.
    
    To show that F2 holds, suppose that there exist agents $a_i\neq a_j\in A$ such that $(a_i\;a_j)\notin\Pi$ but $a_j \succ_i \Pi_r^{-1}(a_i)$ and $a_i \succ_j \Pi_s^{-1}(a_j)$ for some $\Pi_r, \Pi_s\in \Pi$. By construction of $\Pi$, $a_i$ and $a_j$ cannot be fully matched in $(S,P)$, as otherwise we would have $(a_i \; a_j)\in \Pi$. But $\Pi_r^{-1}(a_i) \succeq_i$ worst$_i(P(a_i))$ and $\Pi_s^{-1}(a_j) \succeq_j$ worst$_j(P(a_j))$, so $a_j \succ_i$ worst$_i(P(a_i))$ and $a_i \succ_j$ worst$_j(P(a_j))$, contradicting G2. Thus, $\Pi$ must satisfy F2.
    
    Furthermore, by inspection, G3 implies F3. Finally, by construction, the translation from set functions (where an agent appears at most once in each successor and predecessor set) implies that $\Pi$ satisfies F4. By inspection, it is clear that all permutations in $\Pi$ are distinct (except cycles of length 1).
\end{proof}

\section{Computing a Generalised Stable Partition}
\label{sec:algorithms}

Having shown that the notions of a GSP1 and a GSP2 are equivalent, we will use them interchangeably from hereon and simply refer to a GSP. In most cases, we find it more intuitive to think of a GSP1 rather than a GSP2, but it should be clear from the way in which we refer to the GSP (in terms of its cycles or its sets) which one is used. To show the existence of a GSP and give an upper bound on the computational complexity of finding one, we will show that there is a tight connection to stable half-matchings (in the sense of \citet{Fleiner02, Fleiner08}). Then, we will analyse the complexity of computing a GSP by leveraging an efficient algorithm to compute a stable half-matching in an {\sc sf} instance due to \citet{Fleiner2010}.  
Finally, we will compare this approach to algorithms by \citet{biro10,bmatchingrotations} and \citet{deanmunshi10} for generalisations of {\sc sf} and consider natural greedy approaches.

\subsection{Correspondence with Stable Half-Matchings}

The following relationship allows us to reason about stable half-matchings (stable half-integral $b$-matchings in the language of Fleiner) in the {\sc sf} problem. Note that we already referred to the \emph{half-matching associated with a GSP} in the previous section; here, we will show that this half-matching intuition coincides precisely with the notion of a stable half-matching that was previously studied independently in a different context.

\begin{theorem}
\label{thm:gsphalfcorrespondence}
    Every GSP corresponds to a stable half-matching and vice versa.
\end{theorem}
\begin{proof}
    This can be argued using a mapping between stable half-matchings as edge functions and GSP2. Then, using the previously shown equivalence between GSP1 and GSP2, the result immediately follows. First, let $I=(A,\succ, c)$ be an {\sc sf} instance, let $G$ be its underlying preference graph (where agent preferences over edges are induced from their preferences over the potential partners on the other side of the edge), and let $b$ be the capacity function equivalent to $c$.
    
    For one direction, let $w:E(G)\rightarrow \{0, \frac{1}{2}, 1\}$ be a stable half-matching of $I$. Then by stability of $w$, for all edges $e=\{a_i,a_j\}\in E(G)$, we have that either $w(e)=1$ or there exists $a_k\in e$ such that $\sum_{f : (a_k\in f)\wedge (f\succeq_k e)} w(f) = c_k$. Now let $S,P: A\rightarrow \mathcal{P}(A)$ be set functions such that for all $e=\{a_i,a_j\}\in E(G)$,
    \[
        \begin{dcases*}
            a_i\in S(a_j)\cap P(a_j) \text{ and } a_j\in S(a_i)\cap P(a_i) & if $w(e)=1$; \\
            a_i\in S(a_j) \text{ and } a_j\in P(a_i) & if $w(e)=\frac{1}{2}$ and $\sum_{f : (a_i\in f)\wedge (f\succeq_i e)} w(f) = c_i$; \\
            a_i\in P(a_j) \text{ and } a_j\in S(a_i) & if $w(e)=\frac{1}{2}$ and $\sum_{f : (a_j\in f)\wedge (f\succeq_j e)} w(f) = c_j$; \\
            a_i\notin S(a_j)\cup P(a_j) \text{ and } a_j\notin S(a_i)\cup P(a_j) & if $w(e)=0$.
        \end{dcases*}
    \]
    The differentiation between cases 2 and 3 follows from the definition of the stability of $w$. We deal with agents with free capacity (where $x_i<c_i$) in the same way as in the mapping between GSP1 and GSP2, i.e., we make copies $a_i^1\dots a_i^{c_i-x_i}$ and have $a_i^k\in S(a_i)\cap P(a_i)$ and vice versa to fill the agents to capacity. We claim that, with this construction, $(S,P)$ is a GSP2 for $I$.

    For G1, consider the set differences between $S$ and $P$. Specifically, if there exist some distinct agents $a_i, a_j$ such that $a_j\in S(a_i)\setminus P(a_i)$, then by construction we must have that $a_i\in P(a_j)\setminus S(a_j)$, $w(\{a_i,a_j\})=\frac{1}{2}$ and $\sum_{f : (a_j\in f)\wedge (f\succeq_j \{a_i,a_j\})} w(f) = c_j$. By integrality of $c_j$, there must exist some distinct agent $a_k$ such that $a_k\succ_j a_i$ and $w(\{a_j,a_k\})=\frac{1}{2}$, so it follows that 
    $$\sum_{f : (a_j\in f)\wedge(f\succeq_j \{a_j,a_k\})} w(f) < c_j$$
    and therefore 
    $$\sum_{f : (a_k\in f)\wedge (f\succeq_k \{a_j,a_k\})} w(f) = c_k$$
    by the stability of $w$. This means that $a_k\in S(a_j)$, and it was already established that $a_i\in P(a_j)$. Thus, we obtain the bijection between $P(a_j)$ and $S(a_j)$ with the desired property.

    Suppose that G2 does not hold, i.e., suppose that there exist distinct agents $a_i,a_j$ that are not fully matched to each other such that $a_j$ strictly prefers $a_i$ to their worst predecessor and vice versa. Thus, by construction, we must have either $w(\{a_i,a_j\})=0$ or $w(\{a_i,a_j\})=\frac{1}{2}$. By the stability of $w$, if $w(\{a_i,a_j\})=0$ then we must have either 
    $$\sum_{f : (a_i\in f)\wedge (f\succ_i \{a_i,a_j\})} w(f) = c_i \;\;\text{ or} \sum_{f : (a_j\in f)\wedge (f\succ_j \{a_i,a_j\})} w(f) = c_j.$$
    But then either worst$_i(P(a_i))\succ_ia_j$ or  worst$_j(P(a_j))\succ_j a_i$, respectively, a contradiction. Otherwise, if $w(\{a_i,a_j\})=\frac{1}{2}$ then by the stability of $w$, we must have either 
    $$\sum_{f : (a_i\in f)\wedge (f\succ_i \{a_i,a_j\})} w(f) = c_i - \frac{1}{2} \;\;\text{ or} \sum_{f : (a_j\in f)\wedge (f\succ_j \{a_i,a_j\})} w(f) = c_j - \frac{1}{2}.$$
    But then either $a_j=$ worst$_i(P(a_i))$ or $a_i=$ worst$_j(P(a_j))$, respectively, a contradiction.

    G3 holds by construction, as any half-matching respects the capacity bounds, and any free capacity is filled through the agent copies. G4 also holds by construction, completing this direction of the proof.

    For the other direction, let $(S,P)$ be a GSP2 of $I$ and consider the same mapping as before, i.e., for each edge $e=\{a_i,a_j\}$ (where neither agent is a dummy agent to account for free capacities):
    \[
        w(e) = \begin{dcases*}
            0 & if $a_i\notin S(a_j)\cup P(a_j) \land a_j\notin S(a_i)\cup P(a_j)$; \\
            \tfrac{1}{2} & if $(a_i\in S(a_j) \land a_j\in P(a_i)) \oplus (a_i\in P(a_j) \land a_j\in S(a_i)$); \\
            1 & if $a_i\in S(a_j)\cap P(a_j) \land a_j\in S(a_i)\cap P(a_i)$,
        \end{dcases*}
    \]
    where $\oplus$ denotes exclusive or. We claim that $w$ is a stable half-matching. 

    First, clearly $w$ is a half-matching due to the half-integrality of the mapping and G3 enforcing the capacity constraints. For stability of $w$, consider an edge $e=\{a_i,a_j\}$ where $w(e)\in\{0, \frac{1}{2}\}$, i.e., agents $a_i, a_j$ are not fully matched. Suppose that $w$ is not stable, i.e., 
    $$\sum_{f : (a_i\in f)\wedge (f\succeq_i \{a_i,a_j\})} w(f) < c_i \text{ and} \sum_{f : (a_j\in f)\wedge (f\succeq_j \{a_i,a_j\})} w(f) < c_j.$$ 
    Then $a_j\succ_i$worst$_i(P(a_i))$ and $a_i\succ_j$worst$_j(P(a_j))$, respectively, contradicting G2, so $w$ must be stable as desired. 
\end{proof}

\begin{example}
A simple example to illustrate the correspondence is shown in Tables \ref{table:halfmatching} and \ref{table:gsp2example2}; the former is a stable half-matching (consisting of half-weighted edges in $M^{\text{half}}$ and full-weighted edges in $M^{\text{full}}$) for the instance previously shown in Table \ref{table:unsolvablesf} and the latter is the previously shown GSP2 for the same instance for reference.

\begin{table}[!htb]
\centering
\caption{A half-matching for the instance from Table \ref{table:unsolvablesf}.}
    \begin{tabular}{ c | c  }
    $M^{\text{half}}$ & $M^{\text{full}}$ \\\hline
    $\{\{a_1, a_2\}, \{a_1, a_3\}, \{a_2, a_3\}\}$ & $\{\{a_1, a_4\}, \{a_2, a_4\}, \{a_3, a_5\}\}$
    \end{tabular}
\label{table:halfmatching}
\end{table}

\begin{table}[!htb]
\centering
\caption{A GSP2 for the instance from Table \ref{table:unsolvablesf}.}
    \begin{tabular}{ c | c | c  }
    $a_i$ & $P(a_i)$ & $S(a_i)$ \\\hline
    $a_1$ & $\{ a_3, a_4 \}$ & $\{ a_2, a_4 \}$ \\
    $a_2$ & $\{ a_1, a_4 \}$ & $\{ a_3, a_4 \}$ \\
    $a_3$ & $\{ a_2, a_5 \}$ & $\{ a_1, a_5 \}$ \\
    $a_4$ & $\{ a_1, a_2 \}$ & $\{ a_1, a_2 \}$ \\
    $a_5$ & $\{ a_3 \}$ & $\{ a_3 \}$
\end{tabular}
\label{table:gsp2example2}
\end{table}
\end{example}

This correspondence also immediately gives us the following existence proof for a GSP in {\sc sf}, as we know by Fleiner \citep{Fleiner02} that a stable half-matching always exists, and we now have a mapping to a GSP2, which could also be transformed into a GSP1 using Algorithm \ref{alg:constructgsp1}.

\begin{corollary}
    Any {\sc sf} instance $I$ admits at least one GSP.
\end{corollary}

\subsection{Complexity of Computing Stable Half-Matchings and GSPs in {\sc sf}}
\label{sec:GSPalgorithm}
To prove Theorem \ref{thm:fleinerstructure}, \citet{Fleiner08} showed a one-to-one correspondence between stable half-matchings (in the generalisation of {\sc sf} he studied) and the stable half-matchings in the classical {\sc sr} problem (but allowing incomplete preference lists). This was done using a combination of two different constructions: edge splitting and vertex splitting (on the underlying preference graph of the instance $(G, \succ, b)$).   In the original paper, the correspondence was 
used to establish structural results, and 
no analysis of the time complexity of the transformations was 
presented.  Subsequently, \citet{Fleiner2010} studied stable half-matchings in a many-to-many non-bipartite setting where preference are expressed using \emph{choice functions}.  He showed that, in such a setting where the choice functions satisfy so-called \emph{substitutability} and \emph{increasing} properties, a stable half-matching can be computed in $O(n^2)$ time, where $n$ is the number of agents in the instance, assuming access to constant-time oracles to responses of the choice function. It may be verified that an {\sc sf} instance (possibly with incomplete preference lists) can be modelled using substitutable and increasing choice functions, and that the oracles can be implemented to respond in $O(n)$ time in this setting. The following result therefore follows immediately from Theorem \ref{thm:gsphalfcorrespondence} and the algorithm of \citet{Fleiner2010} to compute a stable half-matching.
\begin{theorem}
\label{theorem:gspcomplexity}
Given an {\sc sf} instance $I$ with $n$ agents, a GSP may be computed in $O(n^3)$ time.
\end{theorem}

Note that the weakly-polynomial inductive and the scaled inductive algorithms for {\sc isa} (a generalization of {\sc sf}) presented by Biró and Fleiner \citep{biro10} can be used to find stable half-matchings in $O(m^2S)$ and $O(m^3\log B)$ time, respectively, where $m$ is the number of edges in the preference graph, $S$ is the sum of all capacities, and $B$ is the largest capacity. In {\sc sf}, given tight bounds $m=O(n^2)$, $S=O(n^2)$ and $B=O(n)$ in the worst case, the algorithms run in polynomial time but only give guarantees of $O(n^6)$ and $O(n^6\log n)$, much worse than the bound we present in Theorem \ref{theorem:gspcomplexity}. 

Interestingly, the Dean-Munshi algorithm \citep{deanmunshi10} to compute a stable allocation in {\sc sa} can also be used to find a stable half-matching in our setting. While an allocation in {\sc sa} need not be half-integral in general, it follows from Lemma 15 in \citep{deanmunshi10} that the Dean-Munshi algorithm does produce a half-integral allocation when all agent and edge capacities are integral. Furthermore, when setting all edge capacities to be 1, the computed half-integral allocation is necessarily a half-matching. Finally, we observe that their definition of stability is equivalent to their definition of stability, so their algorithm indeed returns a stable half-matching when given an {\sc sf} instance. However, it is important to note that their algorithm requires hashing techniques, and therefore results in a time complexity bound of $O(m\log (n))$ time \emph{with high probability}, which translates into an $O(n^2\log (n))$ bound in our setting that applies \emph{with high probability}. This complexity is not guaranteed in general, though, and the algorithm as stated in \citep{deanmunshi10} is a non-deterministic procedure that relies on a probabilistic event to occur in order to terminate. It appears that, although the probability of never terminating monotonically decreases as the instance size increases, any finitely large instance has a non-zero chance of the algorithm not terminating at all. Hence, the Dean-Munshi algorithm can replace the algorithm implicit in Theorem \ref{theorem:gspcomplexity} 
and thereby accelerate the computation of a GSP by an asymptotic factor of $\frac{n}{\log (n)}$ \emph{with high probability}, but it comes with the trade-off of a (generally very small) chance at never terminating.

\subsection{Natural Incremental Algorithms Do Not Work}

It could be expected that, given an {\sc sf} instance, we can compute a GSP incrementally through a series of stable partitions. Specifically, assuming that every agent $a_i$ starts with a capacity $c_i$ of at least 1, we look at the sub-instance in which all agents have capacity 1. We can then compute a stable partition and add its cycles to the GSP. In the next step, we could repeat this and create a new sub-instance where any agent with remaining capacity (i.e., $c_i-1\geq 1$) has capacity 1 and agent pairs already fully matched in our GSP are removed, then compute a stable partition of this instance and add its cycles to the GSP. This could be repeated until either all agents are fully matched in the GSP or there do not exist any mutually acceptable pairs of agents with non-zero capacity in the preference lists.

However, this approach does not work in general -- consider the previously studied instance shown in Table \ref{table:unsolvablesf} that admits the unique stable partition $\Pi'=(a_1 \; a_2 \; a_4)(a_3 \; a_5)$ for the underlying {\sc sr} instance (in the sense described above). However, $\Pi=(a_1 \; a_2 \; a_3)(a_1 \; a_4)(a_2 \; a_4)(a_3 \; a_5)$ is the unique GSP1 of the instance (we will later show that odd-length cycles are invariant and then the uniqueness of these transpositions can be verified by inspection) and clearly $\Pi'\not\subseteq \Pi$ because neither of the cycles in $\Pi'$ is present in $\Pi$.

Similarly, we cannot compute a stable half-matching using this simple incremental logic. Naturally (given the correspondence between stable half-matchings and stable partitions), a stable half-matching $M_1$ (or similarly for an equivalent weight function) for the underlying {\sc sr} instance of Table \ref{table:unsolvablesf} consists of $M_1^{\text{half}}=\{\{a_1,a_2\},\{a_1,a_4\},\{a_2,a_4\}\}$ and $M_1^{\text{full}}=\{\{a_3,a_5\}\}$. The resulting sub-instance with one capacity reduced by one for every agent (and with the preference list entries corresponding to the full match $\{a_3,a_5\}$ removed) is shown in Table \ref{table:wrongsubinstance}. The instance admits the following (unique and integral) stable matching: $M_2=\{\{a_1,a_4\},\{a_2,a_3\}\}$. However, if we put this together with the initial $M_1^{\text{half}}$ and $M_1^{\text{full}}$, we get a half-matching $M_3$ consisting of $M_3^{\text{half}}=\{\{a_1,a_2\},\{a_2,a_4\}\}$ and $M_3^{\text{full}}=\{\{a_1,a_4\}, \{a_2,a_3\}, \{a_3,a_5\}\}$, the resulting instance of which is shown in Table \ref{table:wrongsubinstance2}. Here, we ended up in a situation where agents $a_1,a_4$ each have free capacity of one half but are already fully matched to each other, so they cannot be in another half-match. However, agents $a_2, a_4$ (for example) block $M_3$ with respect to our original {\sc sf} instance shown in Table \ref{table:unsolvablesf} because $a_2$ is half-matched to $a_4$ and $a_1$ but prefers $a_4$ to $a_1$ and $a_4$ still has free capacity of one half which they would rather be matched to $a_2$ (by our assumptions), so this is not a valid approach to arrive at a stable half-matching (which we could then turn into a GSP).

\begin{table}[!htb]
\centering
\begin{minipage}{.49\linewidth}
\centering
\caption{Resulting {\sc sf}/{\sc sr} instance.}
    \begin{tabular}{ c | c | c c c c }
    $a_i$ & $c_i$ & pref \\\hline
    $a_1$ & 1 & $a_2$ & $a_4$ & $a_3$ & $a_5$ \\
    $a_2$ & 1 & $a_4$ & $a_3$ & $a_1$ & $a_5$ \\
    $a_3$ & 1 & $a_1$ & $a_2$ & $a_4$ \\
    $a_4$ & 1 & $a_3$ & $a_1$ & $a_2$ & $a_5$ \\
    $a_5$ & 0 & $a_1$ & $a_2$ & $a_4$ 
    \end{tabular}
\label{table:wrongsubinstance}
\end{minipage}%
\begin{minipage}{.49\linewidth}
\centering
\caption{Resulting instance (2).}
    \begin{tabular}{ c | c | c c c c }
    $a_i$ & $c_i$ & pref \\\hline
    $a_1$ & .5 & $a_2$ & $a_3$ & $a_5$ \\
    $a_2$ & 0 & $a_4$ & $a_1$ & $a_5$ \\
    $a_3$ & 0 & $a_1$ & $a_4$ \\
    $a_4$ & .5 & $a_3$ & $a_2$ & $a_5$ \\
    $a_5$ & 0 & $a_1$ & $a_2$ & $a_4$ 
    \end{tabular}
\label{table:wrongsubinstance2}
\end{minipage} 
\end{table}

\section{Structural Results}
\label{sec:structure}

So far, we have shown how we can generalise the stable partition structure for the {\sc sr} setting to GSPs in the {\sc sf} setting. We have also established a correspondence with stable half-matchings and have given a method to compute a GSP efficiently. In this section, we will establish some more structural properties and implications for the {\sc sf} model.

\subsection{Odd Cycles and Certificate of Unsolvability}
\label{sec:oddcycles}

When generalising the stable partition structure by Tan \citep{tan91_1}, a crucial question is whether odd-length cycles for GSPs are invariant just like in the {\sc sr} setting; the answer is yes, as we now show. Throughout, let alloc$_\Pi(a_i)$ be the number of predecessor-successor pairs in a GSP $\Pi$ that involve $a_i$. We will say that $a_i$ is \emph{allocated} alloc$_\Pi(a_i)$ ``value'' in $\Pi$. 

First, the statement below establishes that the amount an agent is allocated in any stable half-matching (and thus any GSP) is invariant.  This result was established by \citet{Fleiner2010} for stable half-matchings in a many-to-many non-bipartite setting with substitutable and increasing choice functions.  As previously noted in Section \ref{sec:GSPalgorithm}, {\sc sf} is a special case of this general problem model.

\begin{theorem}[\citep{Fleiner2010}]
\label{theorem:fixedpointsfixed}
    Let $I=(A,\succ,c)$ be an {\sc sf} instance and let $w,w'$ be two stable half-matchings of $I$. Then, for all agents $a_i\in A$, it is the case that $\sum_{a_j\in A:j\neq i}w(\{a_i,a_j\})=\sum_{a_j\in A:j\neq i}w'(\{a_i,a_j\})$.
\end{theorem}

This result is similar to the so-called ``Rural Hospitals Theorem'' in the bipartite {\sc Hospitals / Residents} problem \citep{alroth86}, and a weaker analogue for stable matchings in the {\sc sf} problem was established by Irving and Scott \citep{Irving2007}.

\begin{theorem}[\citep{Irving2007}]
    All stable matchings $\mathcal M$ of a solvable {\sc sf} instance $I$ have the same size, and every agent in $I$ has the same number of partners in every $M\in \mathcal M$.
\end{theorem}

Note that it is not necessarily true that an agent that is half-matched in some GSP is half-matched in all GSPs (due to longer cycles of even length that can be broken down into transpositions, as we will establish below). Therefore, it can be the case that an agent has a different number of distinct partners between different GSPs. However, Theorem \ref{theorem:fixedpointsfixed} trivially establishes an analogue of the first part of the theorem above, allowing the following conclusion.

\begin{corollary}
\label{cor:samevalue}
    For every {\sc sf} instance $I=(A,\succ,c)$, every agent is allocated the same value in all GSPs (or stable half-matchings) $\mathcal P$ of $I$.
\end{corollary}

Now to the main result, which establishes the invariance of all odd-length cycles in a GSP.

\begin{theorem}
\label{thm:oddinvariant}
    Let $I$ be an {\sc sf} instance and let $\Pi$ be a GSP1 of $I$. Then the cyclic permutations (also referred to as cycles) of odd length contained in $\Pi$, denoted by $\mathcal{O}_I = \{\Pi_i\in\Pi \; \vert \; \vert A_i\vert \mod 2 = 1 \}$, are invariant, i.e., for any GSP $\Pi'$ of $I$, $\mathcal{O}_I\subseteq \Pi'$.
\end{theorem}
\begin{proof}
    We will differentiate between cycles of length 1 and cycles of odd length at least 3. For both of the arguments, we will leverage the correspondence between stable half-matchings and GSPs and use the mapping from Theorem \ref{thm:gsphalfcorrespondence}. Let $w$ be the stable half-matching associated with $\Pi$.

    For any agent $a_i$, by construction and stability of $\Pi$ it must be that the number of fixed points (or 1-cycles) of $a_i$ is precisely $c_i-\sum_{a_j\in A:j\neq i}w(\{a_i,a_j\})$ and we established in Theorem \ref{theorem:fixedpointsfixed} that $\sum_{a_j\in A:j\neq i}w(\{a_i,a_j\})$ is invariant between all stable half-matchings $w$ of $I$. 

    Now for any cycle (cyclic permutation) $\Pi_i\in \mathcal{O}_I$ with odd length greater than 1, consider $\Pi_i=(a_{i_1} \; a_{i_2} \dots a_{i_k})$, where $k$ is an odd integer greater than 3. By the correspondence established in Theorem \ref{thm:gsphalfcorrespondence}, we have $w(\{a_{i_j},a_{i_{j+1}}\}=\frac{1}{2}$ for all $1\leq j\leq k$ (and addition taken modulo $k$). Thus, using Theorem \ref{thm:fleinerinvariance}, the same will hold in all stable half-matchings of $I$, so again by the correspondence, we must have that $\mathcal{O}_I\subseteq\Pi'$ for all GSP $\Pi'$ of $I$.
\end{proof}

\citet{tan91_2} also showed that for any stable partition in {\sc sr}, any cycle of even length longer than 2 can be broken down into and replaced by a collection of transpositions while maintaining stability. In the following, we show that the analogous result holds for GSPs.

\begin{theorem}
\label{thm:breakcycles}
    Let $I=(A,\succ,c)$ be an {\sc sf} instance and let $\Pi$ be a GSP of $I$. Furthermore, let $\Pi_i=(a_{i_1}\;a_{i_2}\dots a_{i_k})$ be a $k$-cycle (where $k$ is even) in $\Pi$, then $$\Pi'=(\Pi\setminus\Pi_i)\cup(a_{i_1}\;a_{i_2})\dots(a_{i_{k-1}} \; a_{i_k}) \text{ and }\Pi''=(\Pi\setminus\Pi_i)\cup(a_{i_1}\;a_{i_k})\dots(a_{i_{k-2}} \; a_{i_{k-1}})$$
    are GSPs of $I$.
\end{theorem}
\begin{proof}
    We will prove that $\Pi'$ is a GSP of $I$, and the proof for $\Pi''$ follows by symmetry.

    First, notice that the construction is well defined, in the sense that $\Pi'$ consists of distinct (apart from fixed points) cyclic permutations: clearly our construction gives a permutation and note that distinctness would be a problem if one of the newly added transpositions $(a_{i_j}\;a_{i_{j+1}})$ would already be contained in $\Pi$. However, then we would have that $\vert\{s\;\vert\;\Pi_s(a_{i_j})=a_{i_{j+1}}\}\vert + \vert\{s\;\vert\;\Pi_s(a_{i_{j+1}})=a_{i_{j}}\}\vert > 2$ in $\Pi$, violating F4 in Definition \ref{def:gsp1} and thus the stability of $\Pi$. 
    
    Now it is easily verified that $\Pi'$ satisfies F1 because $\Pi$ does, as for all cycles $C\in\Pi\setminus\Pi_i$, we also have $C\in \Pi'$, and all agents $a_s\in A_i$ (recall that $A_i\subseteq A$ is the set of agents involved in the permutation $\Pi_i$) that were previously in $\Pi_i$ are now in transpositions, so their predecessors and successors are equal in this cycle and trivially satisfy F1.

    Similarly, suppose that $\Pi'$ does not satisfy F2, i.e., suppose that there exist distinct agents $a_r,a_s\in A$ such that $(a_r\;a_s)\notin\Pi'$ and we have that $a_s\succ_r (\Pi_u')^{-1}(a_r)$ and $a_r\succ_s (\Pi_v')^{-1}(a_s)$ for some $\Pi_u,\Pi_v\in\Pi'$. However, notice that this could only happen if we had some agents that were in a transposition previously in $\Pi$ that are not in a transposition anymore in $\Pi'$, which is not the case, or at least one agent has a worse predecessor in $\Pi'$ than in $\Pi$. However, the latter case also does not apply, as every cycle except for $\Pi_i$ is still contained in $\Pi'$, i.e., all predecessors are maintained, and every agent in $A_i$ either has their predecessor or their successor (which must be at least as good as their predecessor by F1) from $\Pi_i$ as their predecessor in $\Pi'$ by construction. Thus, no two such agents $a_r$ and $a_s$ can exist, and $\Pi'$ satisfies F2.

    Now for F3, note that for all agents $a_s\in A$, the number of cycles $\Pi_r\in \Pi$ that $a_s$ is contained in must equal $a_s$'s capacity $c_s$ as $\Pi$ satisfies F3, i.e. $\vert\{r\;\vert\; a_s\in A_r\}\vert = c_s$, and, by construction, every agent contained in $\Pi_i$ (if it is contained in $\Pi_i$, then it is contained exactly once in $\Pi_i$ by properties of permutations) is also contained in $(a_{i_1}\;a_{i_2})\dots(a_{i_{k-1}} \; a_{i_k})$ exactly once. Thus, we also have that $\forall a_s\in A$, $\vert\{r\;\vert\; a_s\in A_r'\}\vert = c_s$, where $A_r'$ are the agent sets corresponding to the cyclic permutations in $\Pi'$.

    Finally, for F4, notice that for any two distinct agents $a_u,a_v\in A$, 
    $$\vert\{s\;\vert\;\Pi_s(a_{u})=a_{v}\}\vert + \vert\{s\;\vert\;\Pi_s(a_{v})=a_{u}\}\vert = \vert\{s\;\vert\;\Pi_s'(u)=a_{v}\}\vert + \vert\{s\;\vert\;\Pi_s'(a_{v})=a_{u}\}\vert,$$
    where $\Pi_s'$ are the permutations in $\Pi'$, thus proving that $\Pi'$ satisfies F4, since $\Pi$ satisfies F4.

    Hence, $\Pi'$ satisfies F1-F4 and is a GSP1 (and thus a GSP of $I$)
\end{proof}

\begin{corollary}
\label{cor:reduced}
    Let $I$ be an {\sc sf} instance with $n$ agents, then any non-reduced GSP $\Pi$ of $I$ can be turned into a reduced GSP $\Pi'$ of $I$ in $O(n^2)$ time.
\end{corollary}
\begin{proof}
    Recall that a GSP is reduced if it does not contain any cycles of even length longer than 2. Now each cyclic permutation of odd length will be invariant by Theorem \ref{thm:oddinvariant} and for each cyclic permutation of even length longer than 2, we can decompose it into a collection of transpositions by Theorem \ref{thm:breakcycles}. 

    Notice that $I$ has $n$ agents and each agent has capacity at most $n-1$, so $\Pi$ can contain at most $O(n^2)$ agents in all of its cyclic permutations by F4. Using the method from Theorem \ref{thm:breakcycles}, we can split even cycles in time linear in the number of agents contained in the cycle, so it suffices to do one full scan through $\Pi$ to output $\Pi'$. This establishes the stated complexity.
\end{proof}

Furthermore, the presence of odd cycles rules out the existence of stable matchings, and hence a GSP acts as a suitable succinct certificate for the unsolvability of {\sc sf} instances.

\begin{lemma}
\label{lemma:stablematchingpartition}
    Let $I$ be an {\sc sf} instance and $\Pi$ be a GSP of $I$. If $\Pi$ is reduced and does not contain any cycle of odd length longer than 1, then $\Pi$ corresponds to a stable matching of $I$.
\end{lemma}
\begin{proof}    
    Let $\Pi=(a_{i_1}\;a_{i_2})\dots(a_{i_{2r-1}}\;a_{i_{2r}})(a_{i_{k+1}})\dots(a_{i_{k+l}})$ for $1\leq r\leq k$ where $k\geq 0$ and $l\geq0$, and consider the set of $k$ unordered pairs of agents $M=\{\{a_{i_1},a_{i_2}\},\dots,\{a_{i_{2r-1}},a_{i_{2r}}\}\}$. By properties of GSPs, $M$ is a matching that respects the capacity restrictions of the agents. 

    Now suppose that $M$ is not stable, i.e., there exist distinct agents $a_r,a_s$ such that
    \begin{enumerate}[label=(\roman*)]
        \item  $\{a_r,a_s\}\notin M$, and
        \item $\vert M(a_r)\vert< c_r$ or $a_s\succ_r$ worst$_r(M(a_r))$, and 
        \item $\vert M(a_s)\vert<c_s$ or $a_r \succ_s$ worst$_s(M(a_s))$.
    \end{enumerate}
    Then, by construction of $M$, we have that
    \begin{enumerate}[label=(\roman*)]
        \item  $(a_r\; a_s)\notin \Pi$
        \item
        \begin{enumerate}
            \item either $\vert M(a_r)\vert< c_r$ implies that there exists $\Pi_u=(a_r)\in\Pi$ and, because every agent ranks themselves last, it must be that $\Pi_u=(a_r)\in\Pi$, so $a_s\succ_r\Pi_u^{-1}(a_r)=a_r$, or
            \item $a_s\succ_r$ worst$_r(M(a_r))$ implies that there exists $\Pi_u\in\Pi$ such that $a_s\succ_r\Pi_u^{-1}(a_r)=$ worst$_r(M(a_r))$
        \end{enumerate}
       \item
        \begin{enumerate}
            \item either $\vert M(a_s)\vert< c_s$ implies that there exists $\Pi_v=(a_s)\in\Pi$ and, because every agent ranks themselves last, it must be that $a_r\succ_s\Pi_v^{-1}(a_s)=a_s$, or
            \item $a_r\succ_s$ worst$_s(M(a_s))$ implies that there exists $\Pi_v\in\Pi$ such that $a_r\succ_s\Pi_v^{-1}(a_s)=$ worst$_s(M(a_s))$.
        \end{enumerate}
    \end{enumerate}
    However, then $a_r,a_s$ violate property F2 of Definition \ref{def:gsp1}, contradicting the stability of $\Pi$. Thus, $M$ must be stable.
\end{proof}

\begin{theorem}
\label{thm:certificate}
    $I$ admits a stable matching if and only if no GSP of $I$ contains an odd cycle of length longer than 1.
\end{theorem}
\begin{proof}
    Suppose firstly that $I$ admits a stable matching $M=\{\{a_{i_1},a_{i_2}\},\dots,\{a_{i_{2r-1}},a_{i_{2r}}\}\}$ for $1\leq r\leq k$ where $k\geq 0$. Let $m_i$ be the number of matches (pairs) that agent $a_i$ is contained in within $M$. Now consider the set of permutations $\Pi=(a_{i_1}\;a_{i_2})\dots(a_{i_{2r-1}}\;a_{i_{2r}})(a_{i_{k+1}})\dots(a_{i_{k+l}})$ for $l\geq0$, where the cycles $(a_{i_{k+1}})\dots(a_{i_{k+l}})$ are added in such a way that each agent $a_i$ is in $c_i-m_i$ fixed points, i.e., for every agent $a_i\in A$, $\Pi$ contains $c_i-m_i$ cycles of the form $(a_i)$. This ensures that each agent $a_i$ is contained exactly in $c_i$ cycles. 
    
    By properties of matchings, $\Pi$ contains distinct (apart from fixed points) cyclic permutations as required for a GSP. Furthermore, F1, F3, and F4 are trivially satisfied. Finally, F2 is satisfied because otherwise the blocking agents would also block in $M$. Notice that $\Pi$ does not contain any cycles of odd length longer than 1, and recall that by Theorem \ref{thm:oddinvariant}, odd-length cycles are invariant between GSPs. Thus, no GSP of $I$ contains an odd cycle of length longer than 1.

    Conversely, suppose that no GSP of $I$ contains an odd cycle of length longer than 1. Let $\Pi$ be a GSP of $I$; by Corollary \ref{cor:reduced} we can further assume that $\Pi$ is reduced. Then by Lemma \ref{lemma:stablematchingpartition}, $\Pi$ corresponds to a stable matching $M$ of $I$, so $I$ is solvable as required.
\end{proof}

\begin{corollary}
\label{cor:reducedpartitionmatchingcorrespondence}
    For a solvable {\sc sf} instance $I$, there is a bijective correspondence between the set of stable matchings and the set of reduced GSPs.
\end{corollary}
\begin{proof}
    Notice that the (well-defined) mapping between the stable matchings and reduced GSPs in the proofs of Lemma \ref{lemma:stablematchingpartition} and Theorem \ref{thm:certificate} are inverses of each other and give a bijection.
\end{proof}

Note that Theorems \ref{thm:oddinvariant} and \ref{thm:certificate} together generalise Tan's Theorem \ref{thm:tan}. Finally, we give the following remark concerning the following for the number of GSPs and the complexity of finding \emph{all} GSPs of an instance.

\begin{observation}
Let $I$ be an {\sc sf} instance, let $\mathcal M$ be the set of all stable matchings of $I$, let $\mathcal{RP}$ be the set of reduced GSPs of $I$, and let $\mathcal P$ be the set of GSPs of $I$. Then $\mathcal M\subseteq \mathcal{RP}\subseteq \mathcal P$ (under the natural correspondence between matchings and partitions), and thus $\vert \mathcal{M}\vert\leq\vert \mathcal{RP}\vert\leq \vert \mathcal{P}\vert$. Specifically, because the number of stable matchings can be exponential in the number of agents \citep{gusfield89}, there can also be an exponential number of (reduced and non-reduced) stable partitions, in which case listing each once would take exponential time. However, an efficient enumeration (in the number of reduced or non-reduced stable partitions) could be done using the Fleiner gadget and the stable partition enumeration algorithm due to \citet{glitzner2024structural}.
\end{observation}

It may also interest the reader that the structure of stable half-matchings, and thus GSPs, can also be characterised in terms of the \emph{symmetric stable solutions} in a bipartite many-to-many matching instance derived from a (non-bipartite) {\sc sf} instance by a specific kind of duplication technique studied by \citet{deanmunshi10}.

\subsection{Uniform Capacities and Near-Feasible Stable Matchings}
\label{sec:uniform}

One could think that the existence of a stable half-matching in instances with $c_i=1$ for all $a_i$ (i.e., {\sc sr} instances) implies the existence of a stable matching for instances with $c_i=2$ for all agents $a_i$. However, this is not the case as Table \ref{table:unsolvablesfc2} below shows. The underlying (unsolvable) {\sc sr} instance admits the stable partition $\Pi=(a_1 \; a_2 \; a_3)(a_4 \; a_5)(a_6 \; a_7)(a_8 \; a_9)$. Returning to the case that all agent capacities are equal to 2, by inspection of the preferences, for any stable matching $M$ (if one exists), 
$$\{\{a_1, a_2\}, \{a_1, a_3\}, \{a_2, a_3\}, \{a_4, a_5\}, \{a_6, a_7\}, \{a_8, a_9\} \}\subseteq M.$$ 
However, $a_4,a_6,a_8$ and $a_5,a_7,a_9$ form odd cycles, and hence no stable matching can exist. 

\begin{table}[!htb]
\centering
\caption{An unsolvable {\sc sf} instance with $c_i=2$ and a generalised stable partition indicated using boxes (for transpositions) and solid and broken circles for other successors and predecessors, respectively.}
    \begin{tabular}{ c | c | c c c }
    $a_i$ & $c_i$ & preferences \\\hline
    $a_1$ & 2 & $\boxed{a_2}$ & $\boxed{a_3}$ \\ 
    $a_2$ & 2 & $\boxed{a_3}$ & $\boxed{a_1}$ \\
    $a_3$ & 2 & $\boxed{a_1}$ & $\boxed{a_2}$ \\
    $a_4$ & 2 & $\boxed{a_5}$ & \circled{$a_6$} & \dottedcircled{$a_8$} \\
    $a_5$ & 2 & $\boxed{a_4}$ & \circled{$a_7$} & \dottedcircled{$a_9$} \\
    $a_6$ & 2 & $\boxed{a_7}$ & \circled{$a_8$} & \dottedcircled{$a_4$} \\
    $a_7$ & 2 & $\boxed{a_6}$ & \circled{$a_9$} & \dottedcircled{$a_5$} \\
    $a_8$ & 2 & $\boxed{a_9}$ & \circled{$a_4$} & \dottedcircled{$a_6$} \\
    $a_9$ & 2 & $\boxed{a_8}$ & \circled{$a_5$} & \dottedcircled{$a_7$} 
    \end{tabular}
\label{table:unsolvablesfc2}
\end{table}

Clearly, if $c_i=n-1$ for all agents $a_i$, then the instance is solvable as every agent can be matched to every other agent, and then no two agents can block the matching. However, this is a sharp threshold, as we can construct a family of instances with $k+3$ agents and a uniform capacity function $c_i=k+1$ (for any $k\geq 2$) that is always unsolvable, as illustrated in Table \ref{table:unsolvablesffamily}. Here, any GSP contains the 3-cycle $(a_{k+1} \; a_{k+2} \; a_{k+3})$, independent of the choice of $k$.

\begin{table}[!htb]
\caption{A family of unsolvable {\sc sf} instances with a generalised stable partition indicated using boxes (for transpositions) and solid and broken circles for other successors and predecessors, respectively.}
\centering
    \begin{tabular}{ c | c | c c c c c }
    $a_i$ & $c_i$ & preferences \\\hline
    $a_1$ & $k+1$ & $\boxed{a_{k+1}}$ & $\boxed{a_{k+2}}$ & $\boxed{a_{k+3}}$ & \dots \\ 
    \dots & \dots & \dots \\
    $a_k$ & $k+1$ & $\boxed{a_{k+1}}$ & $\boxed{a_{k+2}}$ & $\boxed{a_{k+3}}$ & \dots \\ 
    $a_{k+1}$ & $k+1$ & $\boxed{a_1}$ & \dots & $\boxed{a_k}$ & \circled{$a_{k+2}$} & \dottedcircled{$a_{k+3}$} \\
    $a_{k+2}$ & $k+1$ & $\boxed{a_1}$ & \dots & $\boxed{a_k}$ & \circled{$a_{k+3}$} & \dottedcircled{$a_{k+1}$} \\
    $a_{k+3}$ & $k+1$ & $\boxed{a_1}$ & \dots & $\boxed{a_k}$ & \circled{$a_{k+1}$} & \dottedcircled{$a_{k+2}$}     
    \end{tabular}
\label{table:unsolvablesffamily}
\end{table}

\begin{observation}
    No uniform capacity function other than $c_i=n-1$ (for all $n$ agents $a_i$) guarantees solvability.
\end{observation}

Thus, it is easy to see and indeed well known that the solvability of preference systems highly depends on the capacities of the agents. Recently, in a paper on hypergraph matching, Csáji \citep{gergely25} provided a polynomial-time algorithm using Scarf's Lemma for the problem of turning an unsolvable {\sc sf} instance $I$ into a solvable {\sc sf} instance $I'$ such that every agent's capacity is modified by at most 1 (up or down) and the sum of all ($\pm 1$) capacity modifications is at most 1. We will provide a simpler algorithm based on GSPs for this problem, give tighter bounds on the time complexity, and analyse the total number of modifications.

Our Algorithm \ref{alg:constructnearfeasible} works as follows: Given an {\sc sf} instance $I=(A,\succ,c)$ and a GSP $\Pi$ of $I$, we decompose all cycles of length longer than 2 into transpositions (and we will argue below how this can be done even for odd-length cycles, despite only having shown so far that even-length cycles can be decomposed into transpositions in Theorem \ref{thm:breakcycles}), turn all transpositions into matches, and alternatively increasing and decreasing the capacity of one agent in each cycle of odd length at least 3 (we denote the set of such cycles by $\mathcal{O}_I^{\geq 3}$ and denote the set of modified agents by $O$). This ensures that we increase the capacity of $\left\lceil\frac{\vert O\vert}{2}\right\rceil$ agents by 1 and decrease the capacity of the remaining agents of $O$ by 1 to arrive at $I'=(A,\succ,c')$.

\begin{algorithm}[hbt]
\renewcommand{\algorithmicrequire}{\textbf{Input:}}
\renewcommand{\algorithmicensure}{\textbf{Output:}}

    \begin{algorithmic}[1]

    \Require{$I=(A,\succ,c)$ : an unsolvable {\sc sf} instance; $\Pi$ : a GSP1 of $I$}
    \Ensure{$c'$ : a new capacity function; $M$ : a stable matching of $I'=(A,\succ,c)$; $O$ : the set of agents with modified capacities in $c'$}

    \State $M \gets \varnothing$
    \State $i \gets 0$
    \State $c'\gets c$
    \State $O\gets \varnothing$

    \For{cycle $\Pi_k\in \Pi$}
        \If{$\vert A_k\vert=2$} \Comment{Add the transpositions as matches}
            \State $M$.add$(\{A_k[0], A_k[1]\})$
        \EndIf

        \State $m\gets\vert A_k\vert$
        
        \If{$m>2$} 
            \If{$m$ is even} \Comment{Decompose longer even-length cycles}
                \State $M$.add$(\{A_k[0], A_k[1]\},\{A_k[2], A_k[3]\}\dots\{A_k[m-2], A_k[m-1]\})$
            \Else 
                \If{$i$ is even} \Comment{Decompose longer odd-length cycles by increasing capacity}
                    \State $M$.add$(\{A_k[0], A_k[1]\},\{A_k[2], A_k[3]\}\dots\{A_k[m-1], A_k[0]\})$
                    \State $c'_{A_k[0]}=c_{A_k[0]}+1$
                    \State $O$.add$(A_k[0])$
                \Else \Comment{Decompose longer odd-length cycles by decreasing capacity}
                    \State $M$.add$(\{A_k[1], A_k[2]\},\{A_k[3], A_k[4]\}\dots\{A_k[m-2], A_k[m-1]\})$
                    \State $c'_{A_k[0]}=c_{A_k[0]}-1$
                    \State $O$.add$(A_k[0])$
                \EndIf
                \State $i \gets i+1$
            \EndIf
        \EndIf
    \EndFor
    \State\Return{$c', M,O$}

    \end{algorithmic}
    \caption{\texttt{NearFeasible}$(I, \Pi)$, constructs a near-feasible solvable instance $I'$ and a corresponding stable matching $M$ from a GSP $\Pi$ of $I$.}
    \label{alg:constructnearfeasible}
\end{algorithm}

To argue the correctness of our approach, the following lemma (similar to Theorem \ref{thm:breakcycles}, but this time dealing with odd-length rather than even-length cycles) is crucial.

\begin{lemma}
\label{lemma:reduceincrease}
    Let $\Pi$ be a GSP for an {\sc sf} instance $I=(A,\succ,c)$ and let $C=(a_{i_1} \; a_{i_2} \dots a_{i_k})\in \Pi$ be a cycle with odd $k\geq 3$. Then, for any agent $a_{i_j}$ with $1\leq j\leq k$, the GSP (where addition and subtraction is taken modulo $k$)
    \begin{itemize}
        \item $\Pi'=(\Pi\setminus C)\cup (a_{i_{j+1}} a_{i_{j+2}})(a_{i_{j+3}} a_{i_{j+4}})\dots (a_{i_{j-2}} a_{i_{j-1}})$ is stable in the instance $I'=(A,\succ,c')$, where $c'_l=c_l$ for all $a_l\in A\setminus \{a_{i_j}\}$ and $c_{i_j}'=c_{i_j}-1$;
        \item $\Pi''=(\Pi\setminus C)\cup (a_{i_{j}} a_{i_{j+1}})(a_{i_{j+2}} a_{i_{j+3}})\dots (a_{i_{j-1}} a_{i_{j}})$ is stable in the instance $I''=(A,\succ,c'')$, where $c''_l=c_l$ for all $a_l\in A\setminus \{a_{i_j}\}$ and $c_{i_j}''=c_{i_j}+1$;
    \end{itemize}
\end{lemma}
\begin{proof}    
    First, we will show that we can increase or decrease capacities in this way without violating the {\sc sf} model bound $0<c_i<n$ for any agent because, for the lower bound, any agent starts with positive capacity in $I$ (by definition of {\sc sf}) and we decrease any agent's capacity by at most 1, and for the upper bound, no agent with capacity $n-1$ can be contained in $C$ and we increase any agent's capacity by at most 1.
    
    Specifically, for the lower bound, decreasing any agent $a_i$'s capacity by 1 leads to a strictly non-negative capacity $c_i\geq0$, and if $a_i$'s capacity becomes zero, then we can simply assume that the agent is removed from the instance (by F3 of Definition \ref{def:gsp1} of a GSP1, if $c_i=0$ then $a_i$ would be present in 0 cycles and, trivially, could never block).

    Now, for the upper bound, suppose first that agent $a_i$ has capacity $c_i=n-1$ and $a_i$ is present in $C$. Notice that by F3 of Definition \ref{def:gsp1}, $a_i$ must be in $n-1$ cycles of $\Pi$. Also, either $a_i$ does not have free capacity, i.e., $\Pi$ does not contain fixed points of $a_i$, but then $a_i$ must be in $n-1$ transpositions because Corollary \ref{cor:notwoconsecutive} established that two agents can never appear consecutively in two separate cycles, contradicting $a_i$'s presence in $C$ (a cycle of length at least 3). Thus, $a_i$ must be in a fixed point in $\Pi$, i.e., must have free capacity. Let $a_p=C^{-1}(a_i)$ be the predecessor of $a_i$ in $C$. By the assumption that every agent ranks themselves last and our assumption that $a_i$ is in a fixed point, we must have that $a_p\succ_i C'^{-1}(a_i)= a_i$ for some cycle $C'=(a_i)$ in $\Pi$. Furthermore, by F1, we must have that $a_i\succeq_p C^{-1}(a_p)$ and by the length of the cycle, we must have that $a_i\neq C^{-1}(a_p)$, therefore $a_i\succ_p C^{-1}(a_p)$. Finally, by F4, we cannot have $(a_i \; a_p)\in\Pi$, therefore F2 is violated and $\Pi$ cannot be stable -- a contradiction. Thus, if $a_i$ is present in $C$, then it must have capacity $c_i<n-1$, so it is possible to increase $a_i$'s capacity by 1 without violating any bounds.
    
    Thus, it is left to show that both partitions are indeed valid GSPs and stable in their respective modified instances. First, we check that $\Pi'$ and $\Pi''$ do indeed satisfy F1, F3, and F4, and then we will argue that they are stable. Notice that F1 is trivially satisfied because we are only replacing one longer cycle by a collection of transpositions which naturally satisfy F1 by containing only two agents each. Furthermore, $a_{i_j}$ is contained in exactly one fewer and one more cycle in $\Pi'$ and $\Pi''$ as in $\Pi$, which corresponds to their changes in capacity, and every other agent is in exactly as many cycles in all three GSPs, therefore F3 is satisfied. Finally, by Theorem \ref{thm:atmostonelonger}, every agent is contained in at most one odd-length cycle, and by Corollary \ref{cor:notwoconsecutive}, no pair of agents can be present consecutively in two separate cycles. Thus, F4 must also be satisfied even after replacing a longer cycle with transpositions.

    Now, with regards to stability, suppose first that $\Pi'$ is not stable in $I'$, i.e., there exist blocking agents $a_r,a_s$ such that $(a_r \; a_s)\notin\Pi'$ but they strictly prefer each other over some predecessor of theirs in $\Pi'$. However, notice that $\Pi$ does not contain any transpositions that are not also in $\Pi'$ in our construction, and also that no agent has a worse predecessor in $\Pi'$ than in $\Pi$. Specifically, for every agent not contained in $C$, they have the exact same successors and predecessors in $\Pi'$ as they do in $\Pi$, and regarding agents in $C$, $a_{i_j}$ has one predecessor and one successor fewer, agents $a_{i_{j+1}}, a_{i_{j+3}}, \dots, a_{i_{j-2}}$ have exactly the same successors but one strictly better predecessor each, and finally agents $a_{i_{j+2}}, a_{i_{j+4}}, \dots, a_{i_{j-1}}$ have the exact same predecessors but one worse successor each. Thus, by stability of $\Pi$ in $I$, $\Pi'$ must be stable in $I'$.
    
    Similarly, for $\Pi''$ in $I''$, every transposition of $\Pi$ is also contained in $\Pi''$. Furthermore, the same as in the argument above for $\Pi'$, no agent has a worse predecessor in $\Pi''$ than in $\Pi$. Thus, by stability of $\Pi$, $\Pi''$ must also be stable.
\end{proof}

We now present the main result.

\begin{theorem}
    Let $I=(A,\succ,c)$ be an unsolvable {\sc sf} instance with $n\geq 3$ agents and let $\Pi$ be a GSP of $I$. Then we can use Algorithm \ref{alg:constructnearfeasible} to find a modified instance $I'=(A,\succ,c')$ and a stable matching $M$ of $I'$ in $O(n^2)$ time ($O(n^3)$ if $\Pi$ is initially unknown)
    such that for all $a_i\in A$, we have that $c_i'\in\{c_i-1, c_i,c_i+1\}$, and $\sum_{a_i\in A}(c_i'-c_i)=\vert\mathcal{O}_I^{\geq3}\vert\mod 2\,\leq 1$. Furthermore, $\sum_{a_i\in A}\vert c_i'-c_i\vert =  \vert\mathcal{O}_I^{\geq3}\vert\leq \frac{n}{3}$, where $\mathcal{O}_I^{\geq3}$ are the cycles of $\Pi$ with odd length at least 3.
\end{theorem}
\begin{proof}
    Our algorithm above is correct because, by Theorem \ref{thm:atmostonelonger}, every agent is contained in at most one odd-length cycle, and by Corollary \ref{cor:notwoconsecutive}, no pair of agents can be present consecutively in two separate cycles. Thus, $O$ consists of $\vert\mathcal{O}_I^{\geq 3}\vert$ distinct agents. Furthermore, for each of these agents, we showed in Lemma \ref{lemma:reduceincrease} that we can decompose the odd-length cycles (of length at least 3) into transpositions when decreasing or increasing the capacity of the agents in $O$ by 1 exactly by the way of the algorithm to arrive at a stable partition without any cycles of odd length at least 3. Thus, by Theorem \ref{thm:certificate}, the resulting instance $I'=(A,\succ,c')$ must be solvable. 
    
    It is easy to see in the pseudo-code that $M$ corresponds exactly to a reduced stable partition (without fixed points) derived by breaking down even-length cycles using Lemmas \ref{thm:breakcycles} and \ref{lemma:reduceincrease}, so by Lemma \ref{lemma:stablematchingpartition}, $M$ is a stable matching of $I'$.

    While $c_i'\in\{c_i-1, c_i, c_i+1\}$ by construction, note that $\sum_{a_i\in A}(c_i'-c_i)=\vert\mathcal{O}_I^{\geq 3}\vert\mod 2\leq 1$ holds because 
    \begin{align*}
        \sum_{a_i\in A}(c_i'-c_i) &=\sum_{a_i\in O}(c_i'-c_i)+\sum_{a_i\notin O}(c_i'-c_i) \\
        &=\sum_{a_i\in O}(c_i'-c_i) & \text{by $c'_i=c_i$ for $a_i\notin O$} \\
        &= \left\lceil\frac{\vert O\vert}{2}\right\rceil(+1)+\left(\vert O\vert -\left\lceil\frac{\vert O\vert}{2}\right\rceil\right)(-1) & \text{by our algorithm} \\
        &= 2\left\lceil\frac{\vert O\vert}{2}\right\rceil - \vert O\vert & \text{simplified} \\
        &= \vert O\vert \mod 2 & \text{simplified} \\
        &= \vert \mathcal{O}^{\geq 3}_I\vert \mod 2 & \text{by construction of $O$}
    \end{align*}

    Similarly, 
    \begin{align*}
        \sum_{a_i\in A}\vert c_i'-c_i\vert &=\sum_{a_i\in O}\vert c_i'-c_i\vert+\sum_{a_i\notin O}\vert c_i'-c_i\vert \\
        &=\sum_{a_i\in O}\vert c_i'-c_i\vert & \text{by $c'_i=c_i$ for $a_i\notin O$} \\
        &= \left\vert\left\lceil\frac{\vert O\vert}{2}\right\rceil(+1)\right\vert+\left\vert\left(\vert O\vert -\left\lceil\frac{\vert O\vert}{2}\right\rceil\right)(-1)\right\vert & \text{by our algorithm} \\
        &= \left\lceil\frac{\vert O\vert}{2}\right\rceil +\vert O\vert -\left\lceil\frac{\vert O\vert}{2}\right\rceil & \text{by $\vert O\vert>0$} \\
        &= \vert O\vert & \text{simplified} \\
        &= \vert \mathcal{O}^{\geq 3}_I\vert  & \text{by construction of $O$}
    \end{align*}

    Thus, it is left to argue why $\vert\mathcal{O}_I^{\geq3}\vert\leq \frac{n}{3}$ holds. Clearly $\mathcal{O}_I^{\geq3}$ contains at most $n$ distinct agent occurrences because, as previously argued using Theorem \ref{thm:atmostonelonger}, every agent is contained in at most one odd-length cycle. Furthermore, by definition, any cycle in $\mathcal{O}_I^{\geq3}$ has a length of at least 3, so the bound follows trivially.
    
    Finally, because we have $O(n)$ cycles and $O(n)$ agents in each cycle, our algorithm runs in $O(n^2)$ time. If $\Pi$ is not given, we can compute it in $O(n^3)$ time as per Theorem \ref{theorem:gspcomplexity} before applying our algorithm. 
\end{proof}

Notice that Lemma \ref{lemma:reduceincrease} allows for the algorithm to be modified easily to only ever increase or only ever decrease capacities instead.

\section{Integer Linear Programs and Experiments}
\label{sec:exp}

Now that we have defined GSPs, proved that they can be computed in polynomial time, and established rich structural results, we will consider optimal solutions and investigate, empirically, properties of random {\sc sf} instances through the lens of GSPs. In {\sc sr}, it has been shown recently that the problems of finding a stable half-matching (or similarly a stable partition) with the maximum number of rank 1  assignments, or, similarly, with the least value of rank $r$ assignments (where $r$ is the \emph{regret} of the instance, defined as the minimum rank over all stable half-matchings where no agent is assigned to anyone they rank worse than $r$), is NP-hard \citep{glitzner24sagt}. Furthermore, a special objective that has received much attention is ``minimum egalitarian cost'', which stands for the minimum sum of ranks of all assignments (for half-matchings, contributions are weighted naturally by $0,\frac{1}{2}$ and 1). Ronn \citep{ronn_srt} established that finding a stable matching with minimum egalitarian cost is NP-hard in {\sc sr}, and Glitzner and Manlove \citep{glitzner24sagt} recently extended this result to stable half-matchings. Thus, the following result holds trivially.

\begin{theorem}
    Let $I$ be an {\sc sf} instance with regret $r$. Finding a stable half-matching (or a GSP) of $I$ with a maximum value of rank 1 assignments, minimum value of rank $r$ assignments, or minimum egalitarian cost, is NP-hard.
\end{theorem}

Motivated by this intractability result, we will give integer linear programming formulations (ILP) for these problems.

\subsection{ILP Models}
\label{sec:ilp}

For our formulation, we will roughly follow the definition of a stable half-matching. Consider an {\sc sf} instance $I=(A,\succ,c)$ and the following model which introduces two variables $h_{ij}$ and $f_{ij}$ for each pair of agents $a_i$ and $a_j$ to model whether they are half-matched (i.e., $h_{ij}=h_{ji}=1$) or fully matched (i.e., $f_{ij}=f_{ji}=1$). This avoids using a non-integer variable that can take values $0,\tfrac{1}{2}$, and 1. We also introduce indicator variables $w_{ij}$ and $w_{ji}$ to express whether $a_i$ is fully assigned up to their capacity $c_i$ with agents at least as good as $a_j$ and vice versa for $a_j$, respectively (we will elaborate on this in the correctness proof). Then, we impose capacity, symmetry, and stability constraints and ensure that the variables are binary. Subject to these constraints, the objective function ensures that any solution corresponds to one with a minimum number of half-matches (i.e., a maximum number of full matches, because everyone is assigned the same value due to Corollary \ref{cor:samevalue}) to ensure that the resulting stable half-matching corresponds automatically to a reduced GSP (rather than any GSP). Given that odd-length cycles are invariant (see Theorem \ref{thm:oddinvariant}), this objective function will only minimise the number of even-length cycles. The objective function and constraints are now given as follows.

{\allowdisplaybreaks
\begin{align} 
\min &\sum_{a_i,a_j\in A}h_{ij} \\
\text{s.t.} \sum_{a_j\in A\setminus\{a_i\}}(\tfrac{1}{2}h_{ij}+f_{ij}) &\leq c_i & \forall a_i\in A \\ 
h_{ij}+f_{ij} &\leq 1 & \forall a_i,a_j\in A \\
h_{ij} &= h_{ji} &\forall a_i,a_j\in A\\
f_{ij} &= f_{ji} &\forall a_i,a_j\in A\\
f_{ij}+w_{ij}+w_{ji}&\geq1 & \forall a_i\neq a_j\in A \\
\sum_{a_k:a_k\succeq_ia_j}(\tfrac{1}{2}h_{ik}+f_{ik}) &\geq c_iw_{ij} &\forall a_i,a_j\in A\\
w_{ij}, h_{ij},f_{ij} &\in\{0,1\} &\forall a_i,a_j\in A
\end{align}}

To transform a solution (given in terms of $h_{ij},f_{ij}$ variables) into a stable half-matching (or equivalently into a GSP), we can use the following simple correspondence: construct the following half-matching (this is given in set notation, but the same construction works for a weight function) consisting of 0.5-weight matches:
$$M^{\text{half}}=\{\{a_i,a_j\}\in A\times A\;\vert\; (i\neq j)\wedge (h_{ij}=1)\},$$
and construct the following full matching consisting of weight-1 matches: 
$$M^{\text{full}}=\{\{a_i,a_j\}\in A\times A\;\vert\; (i\neq j)\wedge (f_{ij}=1)\},$$
and then convert the stable half-matching into a GSP using the method described in the proof of Theorem \ref{thm:gsphalfcorrespondence} if desired. The following result establishes the correctness of our approach.

\begin{theorem}
    The ILP above has a feasible solution $\langle\mathbf{w},\mathbf{h},\mathbf{f}\rangle$ if and only if the associated half-matching $M^{\text{half}}, M^{\text{full}}$ (as constructed above) is stable. 
\end{theorem}
\begin{proof}
    First, let $h_{ij},f_{ij}$ be a feasible solution to the ILP. Naturally, our construction of $M^{\text{half}}, M^{\text{full}}$ leads to a half-matching because constraint (8) ensures that all capacities are respected, because $\sum_{a_j\in A\setminus\{a_i\}}(\tfrac{1}{2}h_{ij}+f_{ij}) \leq c_i$ implies that $\tfrac{1}{2}\vert M^{\text{half}}\vert+\vert M^{\text{full}}\vert\leq c_i$, and constraint (9) ensures that agents can be either half-matched or fully matched, but not both. Now consider the stability constraints: by definition, a half-matching is stable if for all pairs of distinct agents $a_i,a_j$ (or, similarly, for all edges in the graph), either the agents are fully matched (i.e., $\{a_i,a_j\}\in M^{\text{full}}$), or at least one of the agents is filled up to their capacity with agents at least as good as the other agent in the pair. Now notice that:
    
    \begin{align*} 
    & f_{ij}+w_{ij}+w_{ji} \geq 1 &\text{constraint (12)}\\
    \Rightarrow & (f_{ij}=1) \text{ or } (w_{ij}=1) \text{ or } (w_{ji}=1) &\text{by (14)}\\
    \Rightarrow & (\{a_i,a_j\}\in M^{\text{full}}) \text{ or } \left(\sum_{a_k:a_k\succeq_ia_j}(\tfrac{1}{2}h_{ik}+f_{ik})\geq c_i\right) \text{ or }\\&\;\;\;\;\;\;\;\;\;\;\;\;\;\;\;\;\;\;\;\;\;\;\;\;\;\;\;\;\;\;\;\; \left(\sum_{a_k:a_k\succeq_ja_i}(\tfrac{1}{2}h_{jk}+f_{jk})\geq c_j\right)  &\text{by construction and (13)}\\
    \Rightarrow & (\{a_i,a_j\}\in M^{\text{full}}) \text{ or } \left(\sum_{a_k:a_k\succeq_ia_j}(\tfrac{1}{2}h_{ik}+f_{ik})=c_i\right) \text{ or }\\&\;\;\;\;\;\;\;\;\;\;\;\;\;\;\;\;\;\;\;\;\;\;\;\;\;\;\;\;\;\;\;\; \left(\sum_{a_k:a_k\succeq_ja_i}(\tfrac{1}{2}h_{jk}+f_{jk})=c_j\right) &\text{by (8)}
    \end{align*}
    and 
    \begin{align*} 
    & \sum_{a_k:a_k\succeq_ia_j}(\tfrac{1}{2}h_{ik}+f_{ik})=c_i\\
    \Rightarrow & \tfrac{1}{2}\vert\{\{a_i,a_k\}\subseteq M^{\text{half}}\;\vert\;a_k\succeq_ia_j\}\vert+\vert\{\{a_i,a_k\}\subseteq M^{\text{full}}\;\vert\;a_k\succeq_ia_j\}\vert=c_i,
    \end{align*}
    i.e., $a_i$ is filled up to capacity with agents at least as good as $a_j$ (and vice versa for $a_j$ when $w_{ji}=1$). Thus, our set of constraints precisely guarantees our stability condition, and the half-matching made up of $M^{\text{half}}$ and $M^{\text{full}}$ is stable.

    Now, for the other direction, let $M^{\text{half}}, M^{\text{full}}$ be a stable half-matching. On the basis of $M^{\text{half}}$ and $M^{\text{full}}$, assign values to the $\mathbf{h},\mathbf{f}$ and $\mathbf{w}$ variables as follows:
   \begin{align*} 
        h_{ij} &= \begin{dcases*}
            1 & if $\{a_{i},a_j\}\in M^{\text{half}}$; \\
            0 & otherwise.
        \end{dcases*}\\
        f_{ij} &= \begin{dcases*}
            1 & if $\{a_{i},a_j\}\in M^{\text{full}}$; \\
            0 & otherwise.
        \end{dcases*}\\
        w_{ij} &= \begin{dcases*}
            1 & if $\vert M^{\text{full}}_{\succeq j}(a_i)\vert+\tfrac{1}{2}\vert M^{\text{half}}_{\succeq j}(a_i)\vert = c_i$; \\
            0 & otherwise.
        \end{dcases*}
    \end{align*}
    where $M^{\text{half}}(a_i)\subseteq M^{\text{half}}$ is the subset containing only pairs involving $a_i$, and $M_{\succeq j}^{\text{half}}(a_i)\subseteq M^{\text{half}}(a_i)$ is the subset of such pairs where the partner is at least as highly ranked as $a_j$ by $a_i$. $M^{\text{full}}(a_i)$ and $M_{\succeq j}^{\text{full}}(a_i)$ are defined in a similar way from $M^{\text{full}}$. We now establish that constraints (8)-(14) must hold for this choice of variables. 
    
    Indeed, by the definition of a half-matching, we have that for every agent $a_i\in A$, $\tfrac{1}{2}\vert M^{\text{half}}(a_i)\vert+\vert M^{\text{full}}(a_i)\vert\leq c_i$. Now by our construction, $\sum_{a_j\in A\setminus\{a_i\}}(\tfrac{1}{2}h_{ij}+f_{ij})= \tfrac{1}{2}\vert M^{\text{half}}(a_i)\vert+\vert M^{\text{full}}(a_i)\vert$, therefore $\sum_{a_j\in A\setminus\{a_i\}}(\tfrac{1}{2}h_{ij}+f_{ij})\leq c_i$ as desired and (8) must hold.

    Similarly, we know that no two agents can be simultaneously half-matched and fully matched, otherwise they would contradict our matching condition, therefore $M^{\text{half}}(a_i)\cap M^{\text{full}}(a_i)=\emptyset$ and so $h_{ij}+f_{ij}\leq 1$ by construction, i.e., (9) must be satisfied.

    (10) and (11) are satisfied by construction, as $\{a_{i},a_j\}\in M^{\text{half}}\Leftrightarrow \{a_j,a_{i}\}\in M^{\text{half}}$, therefore $h_{ij}=1 \Leftrightarrow h_{ji}=1$, and similarly regarding $M^{\text{full}}$ and $f_{ij}$.

    Now, by stability of our half-matching, for every pair of distinct agents $a_i,a_j$, either $\{a_i,a_j\}\in M^{\text{full}}$, in which case $f_{ij}=1$, or one of the agents is filled up to capacity with agents at least as good as the other, in which case at least one of $w_{ij}$ and $w_{ji}$ must take value 1 by construction, so (12) is satisfied.

    Similarly, by construction $w_{ij}=1$ if and only if $\vert M^{\text{full}}_{\succeq j}(a_i)\vert+\tfrac{1}{2}\vert M^{\text{half}}_{\succeq j}(a_i)\vert = c_i$ (and zero otherwise), and it is always the case that $\sum_{a_k:a_k\succeq_ia_j}(\tfrac{1}{2}h_{ik}+f_{ik}) \geq 0$. Therefore, either (13) is trivially satisfied for $w_{ij}=0$, or it is the case that $w_{ij}=1$ and so $\sum_{a_k:a_k\succeq_ia_j}(\tfrac{1}{2}h_{ik}+f_{ik}) = c_i$, therefore satisfying (13).
    
    Constraint (14) is trivially satisfied by construction.
\end{proof}

As promised at the beginning of this section, we now show how we can adapt our ILP formulation to easily incorporate objectives corresponding to NP-hard optimisation problems. For this, the following simple lookup function (which can be implemented efficiently using an array and a single run through the agents' preferences, for example) will be useful: let $index_i(r)$ denote the index (in our given instance's agent ordering) of the agent that $a_i$ ranks in their $r$'th position (starting from 1, i.e., $index_i(1)$ will be the index of the first choice of $a_i$). For the following formulations, we will always assume constraints (8-14).

To find a stable half-matching with the maximum value (i.e., the total weight of agents, taking into account the intensity given by a full- or half-assignment) of first choice assignments, we can use the following objective (instead of our original objective function):
$$\max \;\sum_{a_i\in A}(f_{i,index_i(1)}+\tfrac{1}{2} h_{i,index_i(1)})$$

Similarly, to minimise the value of agents assigned to someone of minimum regret rank (and none worse), we need a two-step approach. First, we find the minimum regret, i.e., in the first run of the ILP, we solve for the following:
\begin{align*} 
\min\; &r\\
\text{s.t.} \sum_{a_i\in A}\sum_{0< k< n-r}(f_{i,index_i(r+k)}+h_{i,index_i(r+k)})&=0 \\
r&\in\{1,2,\dots,n-1\}
\end{align*}
and then, in the second step, using the optimal objective value $\hat{r}$ for the ILP above, we proceed to solve the following binary ILP:
\begin{align*} 
\min\; \sum_{a_i\in A}(f_{i,index_i(\hat{r})}+\tfrac{1}{2}h_{i,index_i(\hat{r})})\\
\text{s.t.} \sum_{a_i\in A}\sum_{0< k< n-\hat{r}}(f_{i,index_i(\hat{r}+k)}+h_{i,index_i(\hat{r}+k)})&=0
\end{align*}

Lastly, to find a stable half-matching with minimum egalitarian cost (recall that we defined the cost to be the rank-weighted sum of full- and half-assignments), we can use the following natural objective function:
$$\min \sum_{a_i\in A}\sum_{0<k<n}k(f_{i,index_i(k)}+\tfrac{1}{2}h_{i,index_i(k)})$$

Finally, notice that each of the models is \emph{compact}, in the sense that we have $O(n^2)$ variables and constraints, and we need to solve $O(1)$ ILP models for each problem.

\subsection{Solvability and Odd Cycle Properties}

To test our models and get a better intuition for and understanding of random {\sc sf} instances, we will attempt to answer the following questions empirically:

\begin{itemize}
    \item How do changes in the capacity function affect the solvability probability of {\sc sf} instances?
    \item How many odd cycles can we expect in the GSPs for different random {\sc sf} instances?
    \item Do our proposed ILPs perform reasonably well in practice?
\end{itemize}

We generated preferences uniformly at random and considered uniform capacity functions, i.e., every agent within an instance has the same capacity. This links back to, and is motivated by, our observations in Section \ref{sec:uniform}. In the following, we look at results from instances with $2\leq n\leq 32$ (in increments of 2) and uniform capacity functions that are equal to $1,\left\lceil\frac{n-1}{4}\right\rceil,\left\lceil\frac{n-1}{2}\right\rceil,\left\lceil\frac{3n-3}{4}\right\rceil$ and $n-1$. Our working hypothesis was that we would observe a higher likelihood that at least one stable matching exists for every increase in the capacity function, with the intuition being that there is less potential to block. For each $(n,c)$ pair, we computed 1000 random instances. The ILP-based algorithm was implemented in Python using PuLP \citep{pulp} and solved using Gurobi \citep{gurobi} with default settings, and executed on a Windows 10 machine equipped with an Intel i7-13700 CPU and 32 GB RAM and ran over multiple days.

\begin{figure}[!htb]
    \centering
    \begin{tikzpicture}
        \begin{axis}[
            width=.9\textwidth,
            height=6cm,
            ylabel={Ratio},
            ymin=0.6,       
            ymax=1.01,         
            xmin=2,
            xmax=32,
            xlabel={$n$},
            grid=both,
            grid style={dashed, gray!30},
            cycle list name=color list,
            every axis plot/.append style={thick},
            title={Ratio of solvable instances},
            title style={
                yshift=-1.5ex  
            },
            legend cell align={left},
            legend style={
                    at={(0,0)}, 
                    anchor=south west,
                column sep=1ex, 
            },
        ]
            \addplot[acmDarkBlue, mark=*] table [x=n, y=c0] {data/solv.txt};
            \addplot[acmGreen, mark=square*] table [x=n, y=c1] {data/solv.txt};
            \addplot[acmPink, mark=triangle*] table [x=n, y=c2] {data/solv.txt};
            \addplot[acmOrange, mark=diamond*] table [x=n, y=c3] {data/solv.txt};
            \addplot[acmLightBlue, mark=star]  table [x=n, y=c4] {data/solv.txt};
            \addlegendentry{$c=1$}
            \addlegendentry{$c=\left\lceil\frac{n-1}{4}\right\rceil$}
            \addlegendentry{$c=\left\lceil\frac{n-1}{2}\right\rceil$}
            \addlegendentry{$c=\left\lceil\frac{3n-3}{4}\right\rceil$}
            \addlegendentry{$c=n-1$}
        \end{axis}
    \end{tikzpicture}
    \caption{Expected solvability probability of random instances.}
    \label{fig:pn}
    \Description{For all capacity functions except $c=n-1$, the ratio of solvable instances decreases as $n$ increases. For $c=n-1$, the ratio is constant at 1.}
\end{figure}
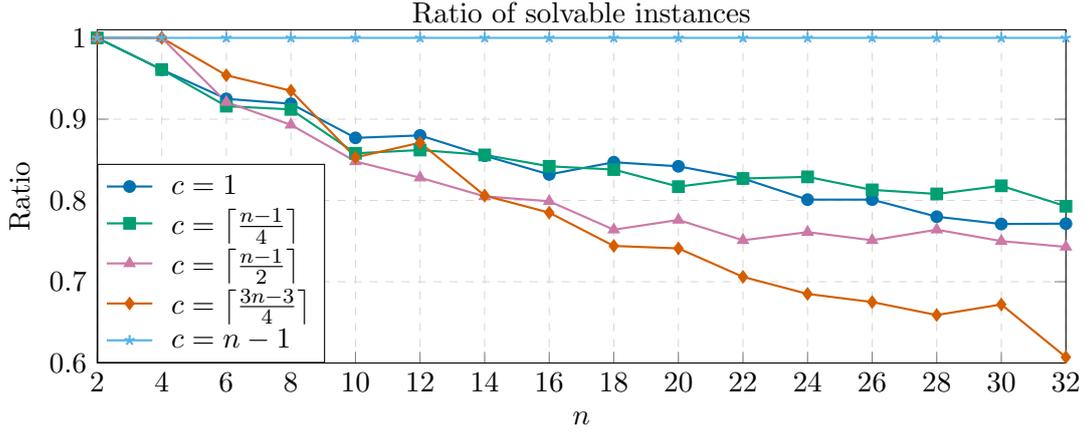

Figure \ref{fig:pn} shows the ratio of solvable instances to all instances for each $(n,c)$ pair. We can see that, for $c=n-1$, there is, of course, always a (trivial) stable matching with every agent being matched to every other agent (by complete preferences). Furthermore, for $c=1$ we recover known results from {\sc sr}, going from 100\% for $n=2$ to around 77\% for $n=32$ (see, for example, \citep{glitzner2024structural,glitzner2025empirics,pittelirving94}). However, what is more interesting is that the data does not support our hypothesis. In fact, although the ratios for all capacity functions (except for $c=n-1$) seem to be very close together for small values of $n$, we can see that for $n\geq 14$, the capacity functions $\left\lceil\frac{n-1}{2}\right\rceil$ and $\left\lceil\frac{3n-3}{4}\right\rceil$ have significantly lower ratios. From a game-theoretic point of view, this means that instances with larger capacities are no more likely, if not strictly less likely, to admit stable matchings than those with smaller capacities (unless the capacities are all $n-1$).

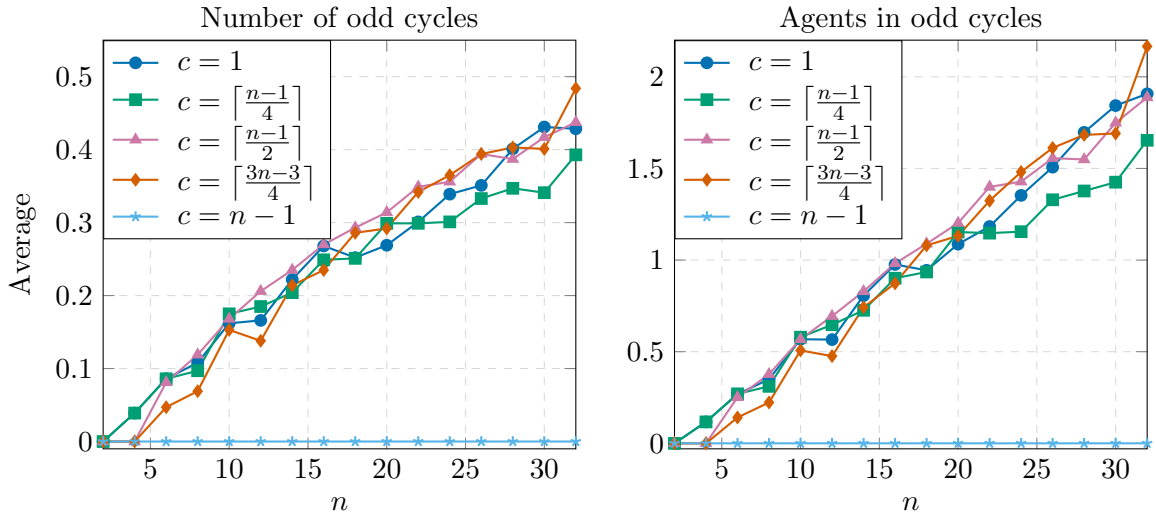
\begin{figure}[!htb]
    \centering

    \begin{subfigure}{0.49\textwidth}
        \begin{tikzpicture}
            \begin{axis}[
                width=\textwidth,
                height=7cm,
                ymin=-0.01,       
                ymax=0.55,         
                xmin=2,
                xmax=32,
                xlabel={$n$},
                grid=both,
                grid style={dashed, gray!30},
                cycle list name=color list,
                every axis plot/.append style={thick},
                ylabel={Average},
                title={Number of odd cycles},
                title style={
                    yshift=-1.5ex  
                },
                legend cell align={left},
                legend style={
                    at={(0,1)}, 
                    anchor=north west,
                    column sep=1ex, 
                },
            ]
                \addplot[acmDarkBlue, mark=*] table [x=n, y=c0] {data/oddcycles.txt};
                \addplot[acmGreen, mark=square*] table [x=n, y=c1] {data/oddcycles.txt};
                \addplot[acmPink, mark=triangle*] table [x=n, y=c2] {data/oddcycles.txt};
                \addplot[acmOrange, mark=diamond*] table [x=n, y=c3] {data/oddcycles.txt};
                \addplot[acmLightBlue, mark=star]  table [x=n, y=c4] {data/oddcycles.txt};
            \addlegendentry{$c=1$}
            \addlegendentry{$c=\left\lceil\frac{n-1}{4}\right\rceil$}
            \addlegendentry{$c=\left\lceil\frac{n-1}{2}\right\rceil$}
            \addlegendentry{$c=\left\lceil\frac{3n-3}{4}\right\rceil$}
            \addlegendentry{$c=n-1$}
            \end{axis}
        \end{tikzpicture}
    \end{subfigure}
    \hfill
    \begin{subfigure}{0.49\textwidth}
        \begin{tikzpicture}
            \begin{axis}[
                width=\textwidth,
                height=7cm,
                ymin=-0.03,       
                ymax=2.2,         
                xmin=2,
                xmax=32,
                xlabel={$n$},
                grid=both,
                grid style={dashed, gray!30},
                cycle list name=color list,
                every axis plot/.append style={thick},
                title={Agents in odd cycles},
                title style={
                    yshift=-1.5ex  
                },
                legend cell align={left},
                legend style={
                    at={(0,1)}, 
                    anchor=north west,
                    column sep=1ex, 
                },
            ]
                \addplot[acmDarkBlue, mark=*] table [x=n, y=c0] {data/oddagents.txt};
                \addplot[acmGreen, mark=square*] table [x=n, y=c1] {data/oddagents.txt};
                \addplot[acmPink, mark=triangle*] table [x=n, y=c2] {data/oddagents.txt};
                \addplot[acmOrange, mark=diamond*] table [x=n, y=c3] {data/oddagents.txt};
                \addplot[acmLightBlue, mark=star]  table [x=n, y=c4] {data/oddagents.txt};
            \addlegendentry{$c=1$}
            \addlegendentry{$c=\left\lceil\frac{n-1}{4}\right\rceil$}
            \addlegendentry{$c=\left\lceil\frac{n-1}{2}\right\rceil$}
            \addlegendentry{$c=\left\lceil\frac{3n-3}{4}\right\rceil$}
            \addlegendentry{$c=n-1$}
            \end{axis}
        \end{tikzpicture}
    \end{subfigure}

    \caption{Properties of cycles of odd length at least 3.}
    \label{fig:oddcycles_subplots}
    \Description{For all capacity functions except $c=n-1$, the average number of odd cycles and the average number of agents in odd cycles increase as $n$ increases. For $c=n-1$, both values are constant at 0 for all values of $n$.}
\end{figure}

Figure \ref{fig:oddcycles_subplots} breaks our observations down into the expected number of cycles of odd length at least 3 and the expected number of agents contained within them. Although the plots show a roughly linear increase for both (except when $c=n-1$), it is important to note that, over all studied $(n,c)$ pairs, the average numbers of odd cycles and agents contained within them were always below 0.5 and 2.2, respectively. For our near-feasible Algorithm \ref{alg:constructnearfeasible}, this means that, for the instances we considered in our experiments, we would change less than one agent's capacity on average.

\begin{figure}[!hbt]
    \centering
    
    \begin{subfigure}{0.49\textwidth}
        \begin{tikzpicture}
            \begin{axis}[
                width=\textwidth,
                height=7cm,
                ymin=0,       
                ymax=6650,         
                xmin=2,
                xmax=32,
                xlabel={$n$},
                grid=both,
                grid style={dashed, gray!30},
                cycle list name=color list,
                every axis plot/.append style={thick},
                ylabel={Avg},
                title={Cost of feasible solution},
                title style={
                    yshift=-1.5ex  
                },
                legend cell align={left},
                legend style={
                    at={(0,1)}, 
                    anchor=north west,
                    column sep=1ex, 
                },
            ]
                \addplot[acmDarkBlue, mark=*] table [x=n, y=c0] {data/coststable.txt};
                \addplot[acmGreen, mark=square*] table [x=n, y=c1] {data/coststable.txt};
                \addplot[acmPink, mark=triangle*] table [x=n, y=c2] {data/coststable.txt};
                \addplot[acmOrange, mark=diamond*] table [x=n, y=c3] {data/coststable.txt};
                \addplot[acmLightBlue, mark=star]  table [x=n, y=c4] {data/coststable.txt};
            \addlegendentry{$c=1$}
            \addlegendentry{$c=\left\lceil\frac{n-1}{4}\right\rceil$}
            \addlegendentry{$c=\left\lceil\frac{n-1}{2}\right\rceil$}
            \addlegendentry{$c=\left\lceil\frac{3n-3}{4}\right\rceil$}
            \addlegendentry{$c=n-1$}
            \end{axis}
        \end{tikzpicture}
    \end{subfigure}
    \hfill
    \begin{subfigure}{0.49\textwidth}
        \begin{tikzpicture}
            \begin{axis}[
                width=\textwidth,
                height=7cm,
                ymin=0,       
                ymax=6650,         
                xmin=2,
                xmax=32,
                xlabel={$n$},
                grid=both,
                grid style={dashed, gray!30},
                cycle list name=color list,
                every axis plot/.append style={thick},
                title={Cost of egalitarian solution},
                title style={
                    yshift=-1.5ex  
                },
                legend cell align={left},
                legend style={
                    at={(0,1)}, 
                    anchor=north west,
                    column sep=1ex, 
                },
            ]
                \addplot[acmDarkBlue, mark=*] table [x=n, y=c0] {data/costegal.txt};
                \addplot[acmGreen, mark=square*] table [x=n, y=c1] {data/costegal.txt};
                \addplot[acmPink, mark=triangle*] table [x=n, y=c2] {data/costegal.txt};
                \addplot[acmOrange, mark=diamond*] table [x=n, y=c3] {data/costegal.txt};
                \addplot[acmLightBlue, mark=star]  table [x=n, y=c4] {data/costegal.txt};
            \addlegendentry{$c=1$}
            \addlegendentry{$c=\left\lceil\frac{n-1}{4}\right\rceil$}
            \addlegendentry{$c=\left\lceil\frac{n-1}{2}\right\rceil$}
            \addlegendentry{$c=\left\lceil\frac{3n-3}{4}\right\rceil$}
            \addlegendentry{$c=n-1$}
            \end{axis}
        \end{tikzpicture}
    \end{subfigure}
    \vspace{0.1cm}
    \begin{subfigure}{0.49\textwidth}
        \begin{tikzpicture}
            \begin{axis}[
                width=\textwidth,
                height=7cm,
                ymin=-0.01,       
                ymax=190,         
                xmin=2,
                xmax=32,
                xlabel={$n$},
                grid=both,
                grid style={dashed, gray!30},
                cycle list name=color list,
                every axis plot/.append style={thick},
                ylabel={Avg (s)},
                title={Time to find a feasible solution},
                title style={
                    yshift=-1.5ex  
                },
                legend cell align={left},
                legend style={
                    at={(0,1)}, 
                    anchor=north west,
                    column sep=1ex, 
                },
            ]
                \addplot[acmDarkBlue, mark=*] table [x=n, y=c0] {data/timingstable.txt};
                \addplot[acmGreen, mark=square*] table [x=n, y=c1] {data/timingstable.txt};
                \addplot[acmPink, mark=triangle*] table [x=n, y=c2] {data/timingstable.txt};
                \addplot[acmOrange, mark=diamond*] table [x=n, y=c3] {data/timingstable.txt};
                \addplot[acmLightBlue, mark=star]  table [x=n, y=c4] {data/timingstable.txt};
            \addlegendentry{$c=1$}
            \addlegendentry{$c=\left\lceil\frac{n-1}{4}\right\rceil$}
            \addlegendentry{$c=\left\lceil\frac{n-1}{2}\right\rceil$}
            \addlegendentry{$c=\left\lceil\frac{3n-3}{4}\right\rceil$}
            \addlegendentry{$c=n-1$}
            \end{axis}
        \end{tikzpicture}
    \end{subfigure}
    \hfill
    \begin{subfigure}{0.49\textwidth}
        \begin{tikzpicture}
            \begin{axis}[
                width=\textwidth,
                height=7cm,
                ymin=-0.01,       
                ymax=190,         
                xmin=2,
                xmax=32,
                xlabel={$n$},
                grid=both,
                grid style={dashed, gray!30},
                cycle list name=color list,
                every axis plot/.append style={thick},
                title={Time to find an egalitarian solution},
                title style={
                    yshift=-1.5ex  
                },
                legend cell align={left},
                legend style={
                    at={(0,1)}, 
                    anchor=north west,
                    column sep=1ex, 
                },
            ]
                \addplot[acmDarkBlue, mark=*] table [x=n, y=c0] {data/timingegal.txt};
                \addplot[acmGreen, mark=square*] table [x=n, y=c1] {data/timingegal.txt};
                \addplot[acmPink, mark=triangle*] table [x=n, y=c2] {data/timingegal.txt};
                \addplot[acmOrange, mark=diamond*] table [x=n, y=c3] {data/timingegal.txt};
                \addplot[acmLightBlue, mark=star]  table [x=n, y=c4] {data/timingegal.txt};
            \addlegendentry{$c=1$}
            \addlegendentry{$c=\left\lceil\frac{n-1}{4}\right\rceil$}
            \addlegendentry{$c=\left\lceil\frac{n-1}{2}\right\rceil$}
            \addlegendentry{$c=\left\lceil\frac{3n-3}{4}\right\rceil$}
            \addlegendentry{$c=n-1$}
            \end{axis}
        \end{tikzpicture}
    \end{subfigure}

    \caption{Comparing the feasible against optimal solutions.}
    \label{fig:comparison_subplots}
    \Description{The average cost of a feasible solution found appears to be very close to the average cost of an egalitarian solution. However, the average time to find an egalitarian solution appears to be shorter than for any feasible solution. Both the average cost and the average time increase with $n$.}
\end{figure}
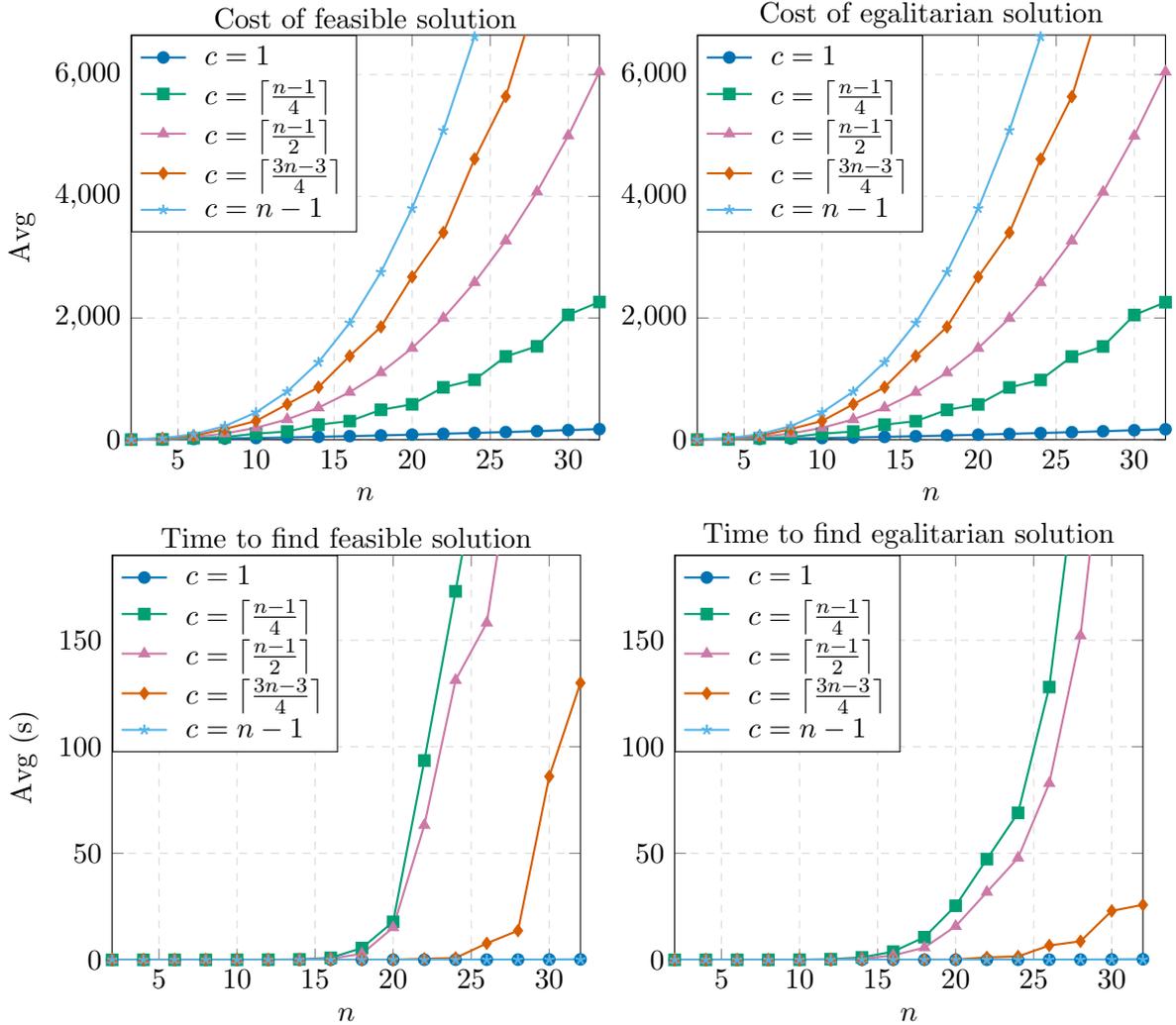

Finally, Figure \ref{fig:comparison_subplots} investigates the differences in egalitarian cost (recall that we defined this to be the sum of ranks in a stable half-matchings, weighted by the intensity of a full- or half-assignment) and computation time when computing a solution using the ILP models from Section \ref{sec:ilp} (only including the time to solve the model and proving optimality, not including model build time). For the plots on the left-hand side, we only used the basic set of constraints (and a trivial objective function minimising $f_{11}$, which is meaningless for our model), and for the plots on the right-hand side, we used the same constraints and the objective function minimising the egalitarian cost of the matching. 

The trend of increasing cost for increasing $n$ and increasing $c$ can be explained easily (larger numbers of both agents and capacities generally increase the total number of matches, thereby increasing the sum of ranks), but it is interesting to see that there is no visible improvement in cost when taking this objective into account. When looking into the data in more detail, we could find some instances where an egalitarian stable half-matching has a slightly better (lower) cost than the stable half-matching computed using the basic model, but the effects are very small on average. 

Surprisingly, the timing results show that the egalitarian model is, on average, faster than the basic model, despite including the time Gurobi takes to prove optimality. The Gurobi logs suggest that the egalitarian objective function helps to guide the solver's search for an integral solution with the help of heuristics, whereas for the basic model, the solver explores many nodes in the search tree before finding an integral solution. One possible explanation for why proving optimality could be reasonably fast is that the number of feasible (stable) solutions could generally be very small, which would align with recent observations for random {\sc sr} instances \citep{glitzner2025empirics}. Another interesting timing-related observation is that while both models are very fast at finding (optimal) solutions to the ILP models for $c=1$ and $c=n-1$, they are increasingly (with growing $n$) very slow for the other capacity functions. This is likely because constraint (13) caused the LP relaxation of these models to be weak.

Of course, when searching for an arbitrary (not necessarily egalitarian) GSP in a large instance, the algorithm underpinning Theorem \ref{theorem:gspcomplexity}
likely terminates much faster in the worst case than the ILP-based implementation. Furthermore, the time required to find an optimal solution to the ILP formulations of NP-hard objectives (such as finding an egalitarian GSP) could likely be improved by first computing an arbitrary GSP
and then fixing the variables corresponding to odd cycles, as they need to be present in every GSP. However, given the small number of agents involved in odd cycles on average that we observed above, the positive effects are likely limited. Furthermore, results for random instances in the {\sc sr} setting suggest that this observation also holds for much larger instances and different classes of random instances \citep{glitzner2025empirics}.

\section{Conclusion}
\label{sec:conclusion}

This work presents a comprehensive structural and algorithmic investigation of the {\sc Stable Fixtures} problem, generalising and significantly extending foundational results on stable partitions from the {\sc Stable Roommates} setting. By introducing and characterising \emph{generalised stable partitions} (GSPs), we provide a unifying framework that captures the structure of non-bipartite matching instances with ordinal preferences and agent capacities. We show that GSPs are efficiently computable, exhibit desirable properties, and offer a certificate of unsolvability as well as a bridge to the theory of stable half-matchings. These results yield new insights into the combinatorial nature of stability in non-bipartite matching markets and strengthen prior work by \citet{tan91_1,tan91_2} and  \citet{fleiner03,Fleiner08}, while connecting to, simplifying, and advancing recent research by \citet{manipulation24}, \citet{gergely25}, and \citet{glitzner2024structural}.

In addition to the structural and algorithmic contributions, we propose a flexible ILP framework for computing optimal stable half-matchings and provide the first empirical study of solution space characteristics in random {\sc sf} instances. We find that the capacity functions seem to have a surprising effect on the likelihood that a random instance is solvable, a phenomenon which could be worthy of further investigation from a theoretical perspective. This work also opens several other avenues for future research. For example, while stability remains central to matching under preference theory, weaker notions such as almost- or semi-stability \citep{abraham06,minimaxprefaamas,herings25} can be useful for settings where integral matchings with some (possibly weakened) stability guarantees are required despite a lack of solvability guarantees. In this direction, \citet{nearfeasibleaamas,nearfeasible} recently unified perspectives on near-feasible stable matchings and almost-stable matchings in a common framework for the {\sc sf} setting. In particular, he considered \emph{optimal near-feasibility} problems that require the minimisation of the total number or the maximum individual amount of capacity modifications required to arrive at a solvable instance, and used the GSP structure that we introduced and studied in this paper to prove that Algorithm \ref{alg:constructnearfeasible} is in fact optimal in the sense that it minimises capacity modifications at the individual and aggregate levels simultanously. Modifications of the algorithm that only ever increase or only ever decrease capacities are also proven to be optimal in this regard.

From an algorithmic perspective, we believe that both GSPs and stable half-matchings can be computed more efficiently, potentially in $O(n^2)$ time, which would improve the practical applicability of these concepts. Another 
interesting direction would be to look at approximation algorithms for optimal {\sc sf} stable matchings and stable half-matchings, e.g., to construct a constant factor approximation algorithm for the many-to-many egalitarian stable (half-)matching problem similar to {\sc sr} \citep{teo97}. We also conjecture that a minimum-regret GSP and stable half-matching can be computed efficiently by proving a bijective correspondence between the set of GSP of an {\sc sf} instance $I$ and the set of stable matchings of a (well-defined) solvable sub-instance $I_T$ of $I$, similar to the result established in the {\sc sr} setting by \citet{glitzner24sagt}, and using the $O(n^2)$ algorithm by \citet{bmatchingrotations} to find a minimum regret stable matching of $I_T$ (adapted from the {\sc sma} setting to the {\sc sf} setting). From an experimental perspective, it would be interesting to implement the $O(n^3)$ algorithm of \citet{Fleiner2010} that underpins Theorem \ref{theorem:gspcomplexity} and compare its runtime to the ILP formulations in Section \ref{sec:ilp} and an implementation of the Dean-Munshi algorithm \citep{deanmunshi10}.
It would also be insightful to extend the experimental study in Section \ref{sec:exp} to more diverse preference distributions, similar to those in \citet{boehmer2024map} and \citet{glitzner2025empirics} in one-to-one settings.

\paragraph{Software and Data Availability} 
Both the software to generate the same instances that we used in our experiments and replicate our results, as well as the raw and processed experimental data, are openly available on Zenodo, see reference \citep{glitzner_2025_15577057}.

\begin{acks}
Frederik Glitzner is supported by a Minerva Scholarship from the School of Computing Science, University of Glasgow. David Manlove was supported by the Engineering and Physical Sciences Research Council, grant number EP/X013618/1. The authors would also like to thank Tam\'as Fleiner for helpful discussions regarding connections between the results of this paper and those of reference \cite{Fleiner2010}, Mathijs Barkel for his support in analysing and interpreting Gurobi logs, 
and the ACM TEAC and SAGT reviewers for their very helpful comments and constructive suggestions. Finally, the authors would like to thank Paul Goldberg, Ron Lavi and Jie Zhang for their support in relation to revisions made to this paper.  For the purpose of open access, the authors have applied a Creative Commons Attribution (CC BY) licence to any Author Accepted Manuscript version arising from this submission.
\end{acks}

\bibliographystyle{ACM-Reference-Format}
\bibliography{papers}


\end{document}